\documentclass[onecolumn,article,asp]{revtex4-1}
\usepackage[latin1]{inputenc}
\usepackage{amsmath}
\usepackage{amsfonts}
\usepackage{amssymb}
\usepackage{graphicx}
\usepackage{textcomp}
\usepackage{subfigure}
\usepackage{geometry}
\usepackage{color}
\begin{document}
\title{Exact diagonalization study of double quantum dots in parallel geometry in zero-bandwidth limit}
\author{Haroon}
\email{haroonjamia@gmail.com}
\author{M.A.H. Ahsan}
\affiliation{Department of Physics, Jamia Millia Islamia (Central University), New Delhi 110025, India}
\begin{abstract}
Exact eigenstates of the parallel coupled double quantum dots attached to the non-interacting leads taken in zero-bandwidth limit are analytically obtained in each particle and spin sector. The ground state of the half-filled system is identified from a four dimensional subspace of the twenty dimensional Hilbert space for different values of tunable parameters of the system viz. the energy levels of the quantum dots, the interdot tunneling matrix-element, the ondot and interdot Coulomb interactions and quantities like spin-spin correlation between the dots, occupancies of the dots are calculated. In the parameter space of the interdot tunneling matrix-element and ondot Coulomb interaction, the dots exhibit both ferromagnetic and antiferromagnetic correlation. There is a critical dependency of the interdot tunneling matrix-element on the ondot Coulomb interaction which leads to transition from the ferromagnetic correlation to the antiferromagnetic correlation as the interdot tunneling matrix-element is increased. The ferromagnetic and antiferromagnetic correlations also exist in the absence of interdot tunneling matrix-element through indirect exchange via the leads. The interdot Coulomb interaction is found to affect this dependency considerably.
\end{abstract}
\maketitle
\section{Introduction}
\label{sysintro}
Recent advances has made it possible to fabricate double quantum dot (DQD) devices {\cite{hjam, jcraig}} at nanoscale. Like single quantum dot, double quantum dots are also highly tunable {\cite{gordon,cronenwett}} and provide a means to examine strongly correlated physics in a controlled setup. In double quantum dots the electrons can be tuned individually and the electronic states in the dots probed in presence of a small tunnel coupling allowed between the dots and the nearby source and drain leads {\cite{mlldg,jjpph}}. A well defined number of electrons also imply a definite confined charge {\it i.e.} integer times the elementary electron charge \cite{cronenwett}. This makes DQD system a leading candidate for the realization of quantum bit {\it i.e.} qubit {\cite{davidp,dloss,dlevs}}. The DQD based devices have applications in quantum information processing, spintronics {\cite{hsds,hgbur}}, spin pumping {\cite{ecragp,blhmrw}} {\it etc}. Double quantum dots provide an ideal model to study interaction between the localized impurity spins in a metal and also between the localized spins and the conduction band electrons. Of the two interesting effects in these systems, the Ruderman-Kittel-Kasuya-Yoshida (RKKY) interaction \cite{ruderman} favors spin ordering and the Kondo effect \cite{hjam, hollitner} causes quenching of individual spins by the conduction electrons. \\
\indent
The probed states in DQDs depend on various parameters such as the geometry of the system, interdot tunneling, coupling with the leads and also the interdot and ondot Coulomb interactions. 
We consider, the DQD system in a parallel geometry and calculate the spin-spin correlation between the dots to explore the nature of probed electronic states. The two quantum dots are identical and tunnel coupled to each other between the connecting non-interacting leads. By considering leads in zero-bandwidth limit, the system becomes a finite site model which can handled using exact diagonalization. The application of zero-bandwidth limit in our model lead it to exactly solvable, analytically. The simple approach used here, has been widely used in quantum dot systems to study transport and magnetic correlation in impurity systems. In earlier studies with zero-bandwidth limit, the results obtained were in good agreement qualitatively, with those obtained experimentally also very approximate to those obtained theoretically with other sophisticated methods {\cite{rallub,racrp}}. With this approach, we have obtained analytical forms of the eigenstates for all possible electron fillings. Using these eigenstates we have calculated spin-spin correlation between the dots, analytically. To understand the behavior of correlation, we have investigated, corresponding occupancies of the dots and the ground state of the system as a function of system parameters.     
\\
\indent
The eigenstate so found are an extension of the isolated DQD system, with the effects of leads incorporated in an approximate way {\cite{bulka}}. The eigenstates thus found in zero-bandwidth limit may be useful in qualitatively understanding transport properties of DQD system \cite{racrp}.
\\ 
\indent
The manuscript is organized as follows: Sec. \ref{modlesystm} contains the description of DQD system in parallel geometry with leads incorporated in zero-bandwidth limit. Analytical results for the system are presented in Sec. \ref{anlticresult}. In Section \ref{analtcspincorr}, we presents analytical calculation of spin-spin correlation in the ground state of half-filled system, Sec. \ref{numbers} contains the numerical results and in the last Sec. \ref{outcome}, we conclude important outcomes.
\section{Model}
\label{modlesystm}
\begin{figure}[h!]
\centering
{\includegraphics[width=0.7\textwidth]{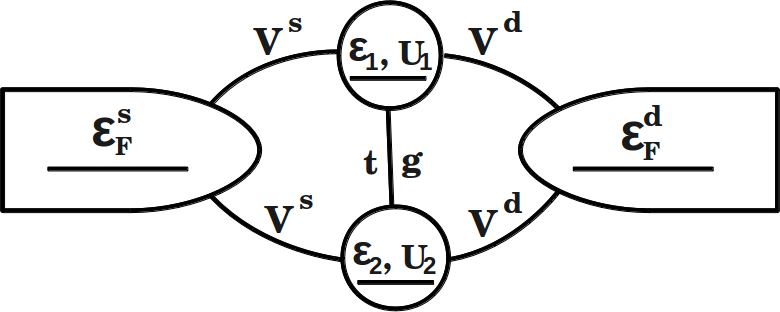}}
\caption{Schematic diagram of DQD system in parallel geometry. The dots are coupled to the source and drain leads through  hybridization parameters $V^{s,d}$. In the ZBW limit, there is only one level$-$the Fermi level$-$ on each of the leads. Dots 1 and 2 are tunnel-coupled through the matrix-element $t$. The dot energies are given by $\varepsilon_{1}$ and $\varepsilon_{2}$ whereas $U_{1}$ and $U_{2}$ are the respective ondot Coulomb interactions. The parameter $g$ denotes the interdot Coulomb interaction.\hfill \break}
\label{zbzmodelparll}
\end{figure}

The system of DQD connected to the leads in parallel geometry is shown in Fig. (\ref{zbzmodelparll}). The system can be described by the two-impurity Anderson model (2IAM) type Hamiltonian \cite{anderson} consisting of three parts
\begin{eqnarray}
{\bf H} = {\bf H}_{dqd} + {\bf H}_{leads} + {\bf H}_{hyb}.
\label{hamilp}
\end{eqnarray}
The Hamiltonian ${\bf H}_{dqd}$ describes the isolated DQD system
\begin{eqnarray*}
{\bf H}_{dqd} &=& \sum_{i=1,2}\varepsilon_{i}\sum_{\sigma}c^{\dagger}_{i\sigma}c_{i\sigma}+ \sum_{i=1,2}U_{i}n_{i\uparrow}n_{i\downarrow}+g\sum_{\sigma,\sigma'}n_{2\sigma}n_{1\sigma'}+
t\sum_{\sigma} \left( c^{\dagger}_{1\sigma}c_{2\sigma} + c^{\dagger}_{2\sigma}c_{1\sigma}\right).
\end{eqnarray*}
The first two terms in ${\bf H}_{dqd}$ represent energies of electrons on spin-degenerate levels $\varepsilon_{i}$ of the dots and $U_{i}$ the respective ondot Coulomb interaction where $i=1,2$ indexes the quantum dots. The third and fourth terms represent the interdot Coulomb interaction $g$ and the interdot tunneling matrix-element $t$ respectively. The Hamiltonian ${\bf H}_{leads}$ describes the source and drain leads where $\varepsilon^{l}_{k\sigma}\ (l=s,d)$ represent the dispersion relation for non-interacting electrons in their continuous energy bands as
\begin{eqnarray*}
{\bf H}_{leads} &=& \sum_{l=s,d}\sum_{k,\sigma}\varepsilon^{l}_{k}c^{\dagger}_{k^{l}\sigma}c_{k^{l}\sigma}.
\end{eqnarray*}
Finally, ${\bf H}_{hyb}$ describes the hybridization of the dots to the leads with possibly spin and {\bf k}-dependent hybridization parameters $V_{ k\sigma}^{l}\ (l=s,d)$
\begin{eqnarray*}
{\bf H}_{hyb} &=& \sum_{l=s,d}\sum_{k,\sigma}\sum_{i=1,2} \left(V^{l}_{k\sigma}c^{\dagger}_{k^{l}\sigma}c_{i\sigma} + V^{l*}_{k\sigma}c^\dagger_{i\sigma}c_{k^{l}\sigma}\right).
\end{eqnarray*}
The zero-bandwidth limit of the leads described by the Hamiltonian ${\bf H}_{leads}$ is taken by replacing the continuous energy band of the leads $\varepsilon^{l}_{k\sigma}\ (l=s,d)$ by the respective Fermi levels with finite degeneracy \cite{racrp,rallub}. The Hamiltonian ${\bf H}$ in eq. (\ref{hamilp}) thus contain a finite number of interacting and non-interacting levels and the problem can managed to be solved exactly. In the present work, we consider only one level, the Fermi level, on both the source and the drain leads \cite{ralcrp}. Thus leads are incorporated in an  approximate way in the DQD system. The model Hamiltonian (\ref{hamilp}) simplifies in zero-bandwidth limit to a four-site Hamiltonian 
\begin{eqnarray}
{\bf H}_{zbw} &=& {\bf H}_{dqd}+ \sum_{l=s,d}\varepsilon^{l}_{F}\sum_{\sigma}c^{\dagger}_{{l}\sigma}c_{{l}\sigma} 
+ \sum_{l=s,d}\sum_{i=1,2}\sum_{\sigma} \left(V^{l}_{i\sigma}c^{\dagger}_{l\sigma}c_{i\sigma} + V^{l*}_{i\sigma}c^\dagger_{i\sigma}c_{l\sigma}\right)
\label{zerobandhamilp}
\end{eqnarray}
where the hybridization parameter $V^{l}_{i\sigma}$ has been assumed to be ${\bf k}$-independent. The DQD system described by  ${\bf H}_{\bf zbw}$ in equation (\ref{zerobandhamilp}) can accommodate up to  $N=8$ electrons and the case $N=4$ electrons correspond to half-filling. The Hamiltonian in eq. (\ref{zerobandhamilp}) is invariant under the source-drain exchange {\it i.e.} symmetric under the transformation $s\leftrightarrow d$ and also invariant under dot-1 and dot-2 exchange {\i.e.} symmetric under the transformation $1\leftrightarrow 2$. 
For the four-site Hamiltonian ${\bf H}_{zbw}$ described in eq. (\ref{zerobandhamilp}) the dimensionality of the Hilbert space is $20\times 20$ in the half-filled case with total spin $S=0$. This difficulty can be further simplified by taking hybridization to the leads $V_{i}^{s\left(s\ast\right)}=V_{i}^{d\left(d\ast\right)}\equiv V_{i}^{\left(\ast\right)}$ to be symmetric  {\cite{dasilva,zitko,ding,lopez}} and the chemical potential in the two leads as $\varepsilon_{F}^{s}=\varepsilon_{F}^{d}\equiv \varepsilon_{F}$, the Fermi energy, corresponding to the equilibrium situation \cite{racrp,bulka}. 
This enables one to transform the fermionic operators of the leads to their symmetric, antisymmetric combinations \cite{rallub} and the Hamiltonian in eq. (\ref{zerobandhamilp}) reduces to a three-plus-one site problem (the antisymmetric combination of lead operators decouples) \cite{rallt}.
\begin{eqnarray}
{\bf H}_{zbw} &=& {\bf H}_{3-site} \oplus {\bf H}_{1-site}
\label{zbw3plus1p}
\end{eqnarray}
Only the symmetric combination of the lead operators $\alpha^{\dag}_{s\sigma}\left|0\right>\equiv \frac{1}{\sqrt{2}}\left(c^{\dag}_{s\sigma}+c^{\dag}_{d\sigma}\right)\left|0\right>$ couples with the dots. The problem now remains to solve the Hamiltonian ${\bf H}_{3-site}$ given as 
\begin{eqnarray}
{\bf H}_{3-site}= {\bf H}_{dqd} + \varepsilon_{F}\sum_{\sigma}\alpha^{\dag}_{s\sigma}\alpha_{s\sigma}
+ \sum_{\sigma}\sum_{i=1,2}\sqrt{2}\left[V_{i\sigma}\alpha^{\dag}_{s\sigma}c_{i\sigma} +V_{i\sigma}^{\ast} c^{\dag}_{i\sigma}\alpha_{s\sigma}\right].
\label{3sitehp}
\end{eqnarray}
We now can find all the eigenvalues of the three site Hamiltonian ${\bf H}_{3-site}$ in eq. (\ref{3sitehp}) for different electron fillings.  The Hamiltonian ${\bf H}_{1-site}$ is the decoupled diagonal term described as 
\begin{eqnarray}
{\bf H}_{1-site}=\varepsilon_{F}\sum_{\sigma}\alpha^{\dag}_{a\sigma}\alpha_{a\sigma}
\label{1sitep}
\end{eqnarray}
where $\alpha^{\dag}_{a\sigma}\left|0\right>\equiv \frac{1}{\sqrt{2}}\left(c^{\dag}_{s\sigma}-c^{\dag}_{d\sigma}\right)\left|0\right>$ is the antisymmetric combination of the fermionic operators in the leads. To obtain any physical quantity for the system we must have eigenenergies and corresponding eigenvectors of the complete four-site Hamiltonian ${\bf H}_{zbw}$. The eigenstates of the Hamiltonian ${\bf H}_{zbw}$ can be constructed by taking direct product of the eigenstates of the Hamiltonians ${\bf H}_{1-site}$ and ${\bf H}_{3-site}$ and the corresponding eigenenergies can also be obtained \cite{allubmag}.
\section{Analytical results}
\label{anlticresult}
The analytical eigenstates of the Hamiltonian ${\bf H}_{3-site}$ have been obtained considering the following simplifications. The hybridization parameters $V_{i\sigma}$ in eq. (\ref{3sitehp}) are taken to be real, symmetric and spin-independent {\it i.e.} $V_{i\sigma}=V_{i\sigma}^{\ast}\equiv V$. The energy levels of the two dots are also taken to be the same {\it i.e.} $\varepsilon_{1}=\varepsilon_{2}\equiv\varepsilon$. Further, for interacting cases, the ondot Coulomb interactions on the two dots are taken to be the same $U_{1}=U_{2}\equiv U$.
\\
\indent
In the absence of ondot and interdot interactions {\it i.e.} $U=0$ and $g=0$, these simplifications allow the transformation of the dot operators to their symmetric and antisymmetric combinations as
$d^{\dag}_{s\sigma}\left|0\right\rangle \equiv \frac{1}{\sqrt{2}}\left(c^{\dag}_{1\sigma}+c^{\dag}_{2\sigma}\right)\left|0\right\rangle$ and $d^{\dag}_{a\sigma}\left|0\right\rangle \equiv \frac{1}{\sqrt{2}}\left(c^{\dag}_{1\sigma}-c^{\dag}_{2\sigma}\right)\left|0\right\rangle$ respectively. Under these transformations the four-site Hamiltonian in eq. ({\ref{zbw3plus1p}}) can be written as 
\begin{eqnarray}
{\bf H}_{zbw}&=&{\bf H}_{eod}\oplus {\bf H}_{da}\oplus {\bf H}_{1-site} 
\label{noninthamil}
\end{eqnarray}
where 
\begin{eqnarray}
{\bf H}_{eod}=\left(\varepsilon+t \right)\sum_{\sigma}d^{\dag}_{s\sigma}d_{s\sigma}+\varepsilon_{F}\sum_{\sigma}\alpha^{\dag}_{s\sigma}\alpha_{s\sigma}+2V\sum_{\sigma}\left(\alpha^{\dag}_{s\sigma}d_{s\sigma}
+d^{\dag}_{s\sigma}\alpha_{s\sigma} \right)
\label{nonint2site}
\end{eqnarray}
is the effective one-dot Hamiltonian with ondot energy and hybridization parameter modified to $\left(\varepsilon+t\right)$ and $2V$ respectively. The Hamiltonian in eq. (\ref{nonint2site}) shows that the symmetric combination of dot-operators $d^{\dag}_{s\sigma}$ couples with the symmetric combination of the leads operators $\alpha^{\dag}_{s\sigma}$. The Hamiltonian ${\bf H}_{da}$ is one-site decoupled diagonal term corresponding to the antisymmetric combination of the dot operators, described as
\begin{eqnarray}
{\bf H}_{da}=\left(\varepsilon-t \right)\sum_{\sigma}d^{\dag}_{a\sigma}d_{a\sigma} \, .
\label{nonint1site}
\end{eqnarray}
The Hamiltonian ${\bf H}_{zbw}$ in eq. (\ref{noninthamil}) is the $(2+1+1)$-site problem and mathematically it is very easy to obtain all eigenvalues and eigenvectors corresponding to all electron numbers.
\subsection{Eigensolutions of ${\bf H}_{1-site}$}
\label{h1siteresult}
For one-electron situation, the eigenstate of ${\bf H}_{1-site}$ is given as ${\bf H}_{1-site}\left|\lambda^{1,0}_{\xi_{0}}\right\rangle=\varepsilon_{F}\left|\lambda^{1,0}_{\xi_{0}}\right\rangle$ where $\left|\lambda^{1,0}_{\xi_{0}}\right\rangle \equiv \alpha^{\dag}_{a\sigma}\left|0\right\rangle$ with $\xi_{0}\equiv\left(N,S,S_{z}\right)=\left(1,\frac{1}{2},+\frac{1}{2}\right)$ and corresponds to the antisymmetric combination of the lead states {\it i.e.} $\alpha^{\dag}_{a\sigma}\left|0\right>=\frac{1}{\sqrt{2}}\left(c^{\dag}_{s\sigma}-c^{\dag}_{d\sigma}\right)\left|0\right>$. The eigenstates are labeled as $\lambda^{1-site,i}_{N,S,S_{z}}$ where the triad $\left( N, S, S_{z} \right)$ is the set of quantum numbers labeling the state in $N$ particle sector; $S$ and $S_{z}$ being the total spin and its $z$-component, respectively.
For two-electron situation the eigenstate is a singlet given as ${\bf H}_{1-site}\left|\lambda^{1,0}_{\xi}\right\rangle=2\varepsilon_{F}\left|\lambda^{1,0}_{\xi}\right\rangle $ with  $ \left|\lambda^{1,0}_{\xi}\right\rangle \equiv \alpha^{\dag}_{a\uparrow}\alpha^{\dag}_{a\downarrow}\left|0\right\rangle$ and 
$\xi=\left(2,0,0 \right)$.
\subsection{One electron eigenstates of ${\bf H}_{3-site}$ Hamiltonian}
\label{h3site1e}
We now consider one-electron solution of ${\bf H}_{3-site}$ in the subspace represented by set of quantum numbers $\xi_{0}\equiv\left( N, S, S_{z} \right)=\left(1,\frac{1}{2},+\frac{1}{2}\right)$.
The one-electron basis states are given as $\left| 1\right\rangle_{1e}=c^{\dag}_{1\sigma}\left|0\right\rangle$, $\left|2\right\rangle_{1e}=c^{\dag}_{2\sigma}\left|0\right\rangle$, $\left|3\right\rangle_{1e}=\alpha^{\dag}_{s\sigma}\left|0\right\rangle$ and the Hamiltonian matrix over the basis becomes
\begin{eqnarray*}
\left(
\begin{array}{lll}
\varepsilon & t & \sqrt{2}V 
\\
t & \varepsilon & \sqrt{2}V 
\\
\sqrt{2}V & \sqrt{2}V & \varepsilon_{F}
\\
\end{array}
\right).
\end{eqnarray*}
In the one-electron situation, as is evident, the ondot and interdot interactions do not play any role.
The eigenvalues of the above one-electron Hamiltonian matrix labeled as
 $\lambda^{{3-site},i}_{N,S,S_{z}}$ are given as $\lambda^{3,1}_{\xi_{0}} = \varepsilon - t $, $\lambda^{3,2}_{\xi_{0}} = \frac{1}{2}\left(\varepsilon+\varepsilon_{F}+t-R^{\prime}\right)$ and $\lambda^{3,3}_{\xi_{0}} = \frac{1}{2}\left(\varepsilon+\varepsilon_{F}+t+R^{\prime}\right)$ where  $R^{\prime}=\sqrt{\left(\varepsilon-\varepsilon_{F}+t\right)^{2}+16V^2}$. 
One of the eigenvectors is given as $\left|\lambda_{\xi_{0}}^{3,1}=\varepsilon -t\right\rangle=\frac{1}{\sqrt{2}}\left(\left|1\right\rangle_{1e}-\left|2\right\rangle_{1e}\right)$ and corresponds to the antisymmetric combination of the dot states. The other two eigenstates labeled by $j=2,3$ are given as $\left|\lambda_{\xi_{0}}^{3,j}\right\rangle_{1e} = C_{\xi_{0}}^{j,1}
\left|1\right\rangle^\prime_{1e}+C_{\xi_{0}}^{j,2}\left|3\right\rangle_{1e}$ with $\left|1\right\rangle^{\prime}_{1e}\equiv\frac{1}{\sqrt{2}}\left(\left|1\right\rangle_{1e}+\left|2\right\rangle_{1e}\right)$. The set of coefficients $C_{\xi_{0}}^{j,i}$ are given as $C_{\xi_{0}}^{j,1}=\pm 2V/{\mathcal{D}_{0j}}$, $C_{\xi_{0}}^{j,2}=\pm \left( \lambda_{\xi_{0}}^{3,j}-\varepsilon_{F}\right)/{\mathcal{D}_{0j}}$ where $\mathcal{D}_{0j}={\sqrt{\left(\lambda_{\xi_{0}}^{3,j}-\varepsilon_{F}\right)^{2}+4V^{2}}}$.
If the Fermi levels of the leads are set at $\varepsilon_{F}=0$ zero and the dot levels below it $ \varepsilon < \varepsilon_{F}$ the ground state in this subspace is given by $\left|\lambda_{\xi_{0}}^{3,2}\right>$.
 In the limit $V\rightarrow 0$, the eigenvalues reduces to $\lambda^{3,1}_{\xi_{0}} = \varepsilon - t $, $\lambda^{3,2}_{\xi_{0}} = \varepsilon_{F} $, $\lambda^{3,3}_{\xi_{0}} = \varepsilon + t $. 
The eigenvalues $\lambda^{3,1}_{\xi_{0}}$ and $\lambda^{3,3}_{\xi_{0}}$ correspond to the eigenstates of isolated DQD system and the eigenvalue $\lambda^{3,2}_{\xi_{0}}$ corresponds to the eigenstates of the decoupled leads.
\subsection{Two electron eigenstates of ${\bf H}_{3-site}$ Hamiltonian}
\label{h3site2e}
\subsubsection{Subspace {$N=2$, $S=0$, $S_{z}=0$}}
\label{h3site2eS0}
This subspace may be labeled as $\xi \equiv \left(N,S,S_{z}\right)=(2,0,0)$. With total spin of the two electron $S=0$ zero, it corresponds to a singlet case. The basis states are spatially symmetric and anti-symmetric in spin. The Hilbert space dimensionality in this subspace is six and the basis states are given as 
$ \left|1\right\rangle_{2e} =  \left|i\right\rangle$, 
$ \left|2\right\rangle_{2e} =  \left|\bar{i}\right\rangle$, 
$ \left|3\right\rangle_{2e} =  \left|\Xi\right\rangle$, 
$ \left|4\right\rangle_{2e} = \frac{1}{{\sqrt{2}}}\left(c^{\dag}_{1\uparrow}\alpha^{\dag}_{s\downarrow}+\alpha^{\dag}_{s\uparrow}c^{\dag}_{1\downarrow}\right)\left|0\right\rangle$, 
$ \left|5\right\rangle_{2e} = \frac{1}{{\sqrt{2}}}\left(c^{\dag}_{2\uparrow}\alpha^{\dag}_{s\downarrow}+\alpha^{\dag}_{s\uparrow}c^{\dag}_{2\downarrow}\right)\left|0\right\rangle$, 
$ \left|6\right\rangle_{2e} = \left|l\right\rangle$. The notations $\left|i\right\rangle$, $\left|\bar{i}\right\rangle$, $\left|\Xi\right\rangle$ and $\left|l\right\rangle$ are defined in \ref{shorthand}. The Hamiltonian matrix over the basis becomes
\begin{eqnarray*}
\left(
\begin{array}{cccccc}
2 \varepsilon+U & 0 & \sqrt{2}t & 2V & 0 & 0
\\
0 & 2 \varepsilon+U & \sqrt{2}t & 0 & 2V & 0
\\
\sqrt{2}t & \sqrt{2}t & 2 \varepsilon+g & \sqrt{2}V & \sqrt{2}V & 0
\\
2V & 0 & \sqrt{2}V & \varepsilon+\varepsilon_{F} & t & 2V
\\
0 & 2V & \sqrt{2}V & t & \varepsilon+\varepsilon_{F} & 2V
\\
0 & 0 & 0 &2V & 2V & 2\varepsilon_{F}
\end{array}
\right).
\end{eqnarray*}
Two of the eigenvalues of the above matrix are given as $\lambda^{3,1}_{\xi}=\frac{1}{2}\left(3 \varepsilon+\varepsilon_{F}+U-t-R\right)$ and\break $\lambda^{3,2}_{\xi}=\frac{1}{2}\left(3 \varepsilon+\varepsilon_{F}+U-t+R\right)$ with their corresponding eigenvectors as \break $\left|\lambda^{3,1}_{\xi}\right>=\frac{\pm 1}{\sqrt{2R\left(R-\Delta\right)}}\left[\left( \Delta-R\right)\left|2\right\rangle^\prime_{2e}+4V\left|5\right\rangle^\prime_{2e}\right]$ and $\left|\lambda^{3,2}_{\xi}\right>=\frac{\pm 1}{\sqrt{2R\left(R+\Delta\right)}}\left[\left( \Delta+R\right)\left|2\right\rangle^\prime_{2e}+4V\left|5\right\rangle^\prime_{2e}\right]$. Where $\Delta=\varepsilon-\varepsilon_{F}+t+U$, 
$R=\sqrt{\Delta^{2}+16V^{2}}$ and the states $\left|2\right\rangle^{\prime}_{2e}=\frac{1}{\sqrt{2}}\left(\left|1\right\rangle_{2e}-\left|2\right\rangle_{2e}\right)$, 
$\left|5\right\rangle^{\prime}_{2e}=\frac{1}{\sqrt{2}} \left(\left|4\right\rangle_{2e}-\left|5\right\rangle_{2e}\right)$. 
The remaining eigenstates can be found by diagonalizing the following $4\times 4$ matrix over the basis 
$\left|3\right\rangle_{2e},\ \left|1^{\prime}\right\rangle_{2e}\equiv\frac{1}{\sqrt{2}}\left(\left|1\right\rangle_{2e}+\left|2\right\rangle_{2e}\right),\ \left|4\right\rangle^{\prime}_{2e}\equiv\frac{1}{\sqrt{2}}\left(\left|4\right\rangle_{2e}+\left|5\right\rangle_{2e}\right)$ and $\left|6\right\rangle_{2e}$:
\begin{eqnarray*}
\left(
\begin{array}{llll}
\varepsilon_{1}^{\prime} & 2t & 2V & 0
\\
2t & \varepsilon_{2}^{\prime} & 2V & 0
\\
2V & 2V & \varepsilon_{3}^{\prime} & 2\sqrt{2}V
\\
0 & 0 & 2\sqrt{2}V & 2\varepsilon_{F}
\end{array}
\right)
\end{eqnarray*}
where $\varepsilon_{1}^{\prime}=2\varepsilon+g$, $\varepsilon_{2}^{\prime}=2\varepsilon+U$, $\varepsilon_{3}^{\prime}=\varepsilon+\varepsilon_{F}+t$. The four eigenvalues labeled as $i=3,4,5,6$ are obtained as 
$\lambda^{3,i}_{\xi}=y_{i}-\frac{\mathtt{a_{1}}}{4}$ where 
$\mathtt{a_{1}} =  -\left(\varepsilon_{1}^{\prime}+\varepsilon_{2}^{\prime}+\varepsilon_{3}^{\prime}+2\varepsilon_{F}\right)$.
The values $y_{i}$ are given by $y_{3,4}=\frac{1}{2}\left(-k_{1}\pm \sqrt{k_{1}^2-4m_{1}}\right)$ and 
$y_{5,6}=\frac{1}{2}\left(+k_{1}\pm \sqrt{k_{1}^2-4n_{1}}\right)$. Where $k_{1}$, $m_{1}$ and $n_{1}$ can be obtained from \ref{Biquadsoln} using parameters $\mathtt{b_{1}} = \left(2\varepsilon_{F}+\varepsilon_{1}^{\prime} \right)\left(\varepsilon_{1}^{\prime}+\varepsilon_{2}^{\prime}+\varepsilon_{3}^{\prime} \right)+\varepsilon_{2}^{\prime}\varepsilon_{3}^{\prime}-\varepsilon_{1}^{\prime 2}-4\left(t^{2}+4V^{2} \right)$, 
$\mathtt{c_{1}} = \left(2\varepsilon_{F}+\varepsilon_{3}^{\prime} \right)\left(4t^{2}
-\varepsilon_{1}^{\prime}\varepsilon_{2}^{\prime}\right) +2 \left(\varepsilon_{1}^{\prime}+\varepsilon_{2}^{\prime}\right)\left(6V^{2}-\varepsilon_{F}\varepsilon_{3}^{\prime} \right) +16V^{2}\left(\varepsilon_{F}-t \right)$ and 
$\mathtt{d_{1}} = \left(\varepsilon_{1}^{\prime}\varepsilon_{2}^{\prime}-4t^{2} \right)\left(2\varepsilon_{F}\varepsilon_{3}^{\prime}-8V^{2} \right)-8V^{2}\varepsilon_{F}\left(\varepsilon_{1}^{\prime}+\varepsilon_{2}^{\prime}-4t \right)$. 
The corresponding normalized eigenvectors $\left|\lambda_{\xi}^{3,i}\right\rangle$ 
are given as 
\begin{eqnarray}
\left|\lambda_{\xi}^{3,i}\right\rangle=C^{1,i}_{\xi}\left|3\right\rangle_{2e}+C^{2,i}_{\xi}\left|1\right\rangle^{\prime}_{2e}+C^{3,i}_{\xi}\left|4\right\rangle^{\prime}_{2e}+C^{4,i}_{\xi}\left|6\right\rangle_{2e}. 
\label{gseta1}
\end{eqnarray}
The coefficients $C^{j,i}_{\xi}$ are given by 
$C^{1,i}_{\xi} = \mp \left(2V\gamma_{1i}\gamma_{4i}\right)/\mathcal{D}_{i}$, 
$C^{2,i}_{\xi} = \mp \left(2V\gamma_{2i}\gamma_{4i}\right)/\mathcal{D}_{i}$, 
$C^{3,i}_{\xi} = \pm \gamma_{4i}\gamma_{5i}/\mathcal{D}_{i}$, 
$C^{4,i}_{\xi} = \mp \left(2\sqrt{2}V\gamma_{5i}\right)/\mathcal{D}_{i}$ with 
$\gamma_{1i}=\varepsilon_{2}^{\prime}-2t-\lambda_{\xi}^{3,i}$, 
$\gamma_{2i}=\varepsilon_{1}^{\prime}-2t-\lambda_{\xi}^{3,i}$, 
$\gamma_{3i}=\varepsilon_{1}^{\prime}-\lambda_{\xi}^{3,i}$, 
$\gamma_{4i}=2\varepsilon_{F}-\lambda_{\xi}^{3,i}$, 
$\gamma_{5i}=\left( \gamma_{3i}\gamma_{1i}+2t\gamma_{2i}\right)$ and 
$\mathcal{D}_{i} = \sqrt{\gamma_{5i}^{2}\left(\gamma_{4i}^{2}+8V^{2} \right)+4V^{2}\gamma_{4i}^{2}\left(\gamma_{1i}^{2}+\gamma_{2i}^{2} \right)}$. 
In this subspace the ground state of the two-electron system is given by $\left|\lambda_{\xi}^{3,3}\right\rangle$. 
In the limit $V\rightarrow 0$ with $\varepsilon_{F}=0$, one can obtain the eigenvalues of the isolated DQDs as $\lambda^{1}_{dqd} = 2\varepsilon+U $, $\lambda^{2}_{dqd} = \frac{1}{2}\left[4\varepsilon+g+U-\sqrt{\left(U-g\right)^{2}+16t^{2}} \right] $ and $\lambda^{3}_{dqd} = \frac{1}{2}\left[4\varepsilon+g+U+\sqrt{\left(U-g\right)^{2}+16t^{2}} \right]$. The remaining eigenvalues $\lambda^{3,1}_{\xi} = \varepsilon - t $,  $\lambda^{3,5}_{\xi} =\varepsilon+t$ and $\lambda^{3,6}_{\xi} =0$ appear due to increased dimensionality of the Hilbert space due to  hybridization with the leads. 
\subsubsection{Subspace {$N=2$, $S=1$, $S_{z}=1$}}
\label{h3site2eS1}
This subspace may be labeled as $\xi^{\prime}\equiv \left(N,S,S_{z}\right)=(2,1,1)$. With total spin of two-electrons equal to one, it corresponds to triplet case. For the non-magnetic case it is sufficient to consider the subspace $S_{z}=1$. The basis states are symmetric in spin and antisymmetric with respect to their spatial indices. The Hilbert space is just three dimensional with basis states 
$\displaystyle{\left|7\right\rangle_{2e}=\left|\sigma\right\rangle}$, 
$\displaystyle{\left|8\right\rangle_{2e}=c^{\dag}_{1\uparrow}\alpha^{\dag}_{s\uparrow}\left|0\right\rangle}$ and
$\displaystyle{\left|9\right\rangle_{2e}=c^{\dag}_{2\uparrow}\alpha^{\dag}_{s\uparrow}\left|0\right\rangle}$.
The Hamiltonian matrix over the basis is given as 
\begin{eqnarray*}
\left(
\begin{array}{ccc}
2\varepsilon+g & \sqrt{2}V & -\sqrt{2}V
\\
\sqrt{2}V & \varepsilon+\varepsilon_{F} & t
\\
-\sqrt{2}V& t & \varepsilon+\varepsilon_{F}
\end{array}
\right).
\end{eqnarray*}
The eigenvalues of the above $3\times 3$ matrix can be easily obtained by performing the transformation of the states $\displaystyle{\left|8\right\rangle_{2e}}$ and $\displaystyle{\left|9\right\rangle_{2e}}$ as $\left|8\right>^\prime_{2e}=\frac{1}{\sqrt{2}}\left(\left|8\right\rangle_{2e}-\left|9\right\rangle_{2e}\right)$ and $\left|9\right>^\prime_{2e}=\frac{1}{\sqrt{2}}\left(\left|8\right\rangle_{2e}+\left|9\right\rangle_{2e}\right)$.
 One obtains $\lambda_{\xi^{\prime}}^{3,7}=\varepsilon+\varepsilon_{F}+t$, 
 $\lambda_{\xi^{\prime}}^{3,8}=\frac{1}{2}\left(3\varepsilon+g+\varepsilon_{F}-t
+R_{0}\right)$ and $\lambda_{\xi^{\prime}}^{3,9}=\frac{1}{2}\left(3\varepsilon+g+\varepsilon_{F}-t
-R_{0}\right)$ where $R_{0}=\sqrt{\Delta_{0}^{2}+16V^{2}}$ with $\Delta_{0}=\varepsilon-\varepsilon_{F}+g+t$. 
The one of the eigenvectors is given as $\left|\lambda_{\xi^{\prime}}^{3,7}\right>=\left|9\right>^\prime_{2e}$ and the other two labeled by $j=8,9$ are given as $\left|\lambda_{\xi^{\prime}}^{3,j}\right>=\frac{\pm 1}{\sqrt{2R_{0}\left(R_{0}+\alpha^{j}\Delta_{0}\right)}}\left[-4V\left|7\right\rangle_{2e}+\left(\Delta_{0}+\alpha^{j} R_{0}\right)\left|8\right\rangle^\prime_{2e}\right]$ 
where $\alpha^{j}=-1$ for $j=8$ and $\alpha^{j}=+1$ for $j=9$. If the Fermi levels of the leads are set at $\varepsilon_{F}=0$ zero and the dot levels below it $ \varepsilon < \varepsilon_{F}$ the ground state in this subspace is given by $\left|\lambda_{\xi^{\prime}}^{3,8}\right>$
\subsection{Three electron eigenstates of ${\bf H}_{3-site}$ Hamiltonian }
\label{h3site3eS}
\subsubsection{Subspace $N=3$, $S=1/2$, $S_{z}=1/2$}
\label{h3site3eS1by2}
This subspace may be labeled by the triad $ \xi_{1}\equiv\left(N,S,S_{z}\right)=\left(3,\frac{1}{2},+\frac{1}{2}\right)$. The three electron Hilbert space in the $S^{2}$-symmetry adapted basis with total spin $S=\frac{1}{2}$ and $S_{z}=+\frac{1}{2}$ is eight dimensional. The basis states are given as 
$\left|1\right\rangle_{3e} = \alpha^{\dag}_{s\uparrow}\left|\Xi\right\rangle$ , 
$\left|2\right\rangle_{3e}=\frac{1}{\sqrt{3}}\left(\alpha^{\dag}_{s\uparrow}\left|\Theta \right\rangle-\sqrt{2}\alpha^{\dag}_{s\downarrow} \left|\sigma\right\rangle\right)$, 
$\left|3\right\rangle_{3e}=c^{\dag}_{\bar{i}\uparrow}\left|i\right\rangle$, 
$\left|4\right\rangle_{3e}=\alpha^{\dag}_{s\uparrow}\left|i\right\rangle$, 
$\left|5\right\rangle_{3e}=c^{\dag}_{i\uparrow}\left|\bar{i}\right\rangle$, 
$\left|6\right\rangle_{3e}=\alpha^{\dag}_{s\uparrow}\left|\bar{i}\right\rangle$, 
$\left|7\right\rangle_{3e}=c^{\dag}_{i\uparrow}\left|l\right\rangle$, 
$\left|8\right\rangle_{3e}=c^{\dag}_{\bar{i}\uparrow}\left|l\right\rangle$. Where $\left|\Xi\right\rangle$, $\left|\Theta\right\rangle$, $\left|\sigma\right\rangle$, $\left|i\right\rangle$, $\left|\bar{i}\right\rangle$ and $\left|l\right\rangle$ are defined in {\ref{shorthand}}. The Hamiltonian matrix over the basis is given as
\begin{eqnarray*}
\left(
\begin{array}{cccccccc}
a_{1}^{d} & 0 & -V & \sqrt{2}t & -V & \sqrt{2}t & -V & -V 
\\
0 & a_{1}^{d} & \sqrt{3}V & 0 & -\sqrt{3} V & 0 & -\sqrt{3} V &\sqrt{3} V 
\\
-V & \sqrt{3}V & \varepsilon_{2}^{\prime\prime} &  \sqrt{2}V & -t & 0 & 0 & 0 
\\
\sqrt{2}t & 0 & \sqrt{2}V & a_{3}^{d} & 0 & 0 & -\sqrt{2}V & 0 
\\
- V & -\sqrt{3}V & -t & 0 & \varepsilon_{2}^{\prime\prime} & \sqrt{2}V & 0 & 0 
\\
\sqrt{2}t & 0 & 0 & 0 & \sqrt{2}V & a_{3}^{d} & 0 & -\sqrt{2}V 
\\
-V & -\sqrt{3}V & 0 & -\sqrt{2}V & 0 & 0 & \varepsilon_{4}^{\prime\prime} & t 
\\
-V & \sqrt{3}V & 0 & 0 & 0 & -\sqrt{2}V & t & \varepsilon_{4}^{\prime\prime} 
\end{array}
\right)
\end{eqnarray*}
where $\varepsilon_{2}^{\prime\prime}=3\varepsilon+U+2g$, $\varepsilon_{4}^{\prime\prime}=\varepsilon+2\varepsilon_{F}$. The above matrix under the basis transformation $ \left|3\right\rangle^{\prime}_{3e}=\frac{1}{\sqrt{2}}\left(\left|3\right\rangle_{3e}+\left|5\right\rangle_{3e} \right)$, $ \left|4\right\rangle^{\prime}_{3e}=\frac{1}{\sqrt{2}}\left(\left|4\right\rangle_{3e}+\left|6\right\rangle_{3e} \right)$, $ \left|5\right\rangle^{\prime}_{3e}=\frac{1}{\sqrt{2}}\left(\left|3\right\rangle_{3e}-\left|5\right\rangle_{3e} \right)$, $ \left|6\right\rangle^{\prime}_{3e}=\frac{1}{\sqrt{2}}\left(\left|4\right\rangle_{3e}-\left|6\right\rangle_{3e} \right)$, $ \left|7\right\rangle^{\prime}_{3e}=\frac{1}{\sqrt{2}}\left(\left|7\right\rangle_{3e}+\left|8\right\rangle_{3e} \right)$ and $ \left|8\right\rangle^{\prime}_{3e}=\frac{1}{\sqrt{2}}\left(\left|7\right\rangle_{3e}-\left|8\right\rangle_{3e} \right)$, block diagonalizes into two $4\times 4$ matrices of which one defined over the basis $\left\lbrace \left|1\right\rangle_{3e}, \left|3\right\rangle^{\prime}_{3e}, \left|4\right\rangle^{\prime}_{3e}, \left|7\right\rangle^{\prime}_{3e} \right\rbrace$ is given as 
\begin{eqnarray*}
\left(
\begin{array}{cccc}
a_{1}^{d} & -\sqrt{2}V & 2t & -\sqrt{2}V 
\\
-\sqrt{2}V & a_{2}^{d} & \sqrt{2}V & 0 
\\
2t & \sqrt{2}V & a_{3}^{d} & -\sqrt{2}V 
\\
-\sqrt{2}V & 0 & -\sqrt{2}V & a_{4}^{d} 
\\
\end{array}
\right)
\end{eqnarray*}
where $a_{1}^{d} = 2\varepsilon +\varepsilon_{F}+g$, $a_{2}^{d}=3\varepsilon -t+U+2g$, $a_{3}^{d}=2\varepsilon + U+\varepsilon_{F}$ and $a_{4}^{d}=\varepsilon + t+2\varepsilon_{F}$. The eigenvalues of the above matrix labeled as $i=1,2,3,4$ are given as  $\lambda^{3,i}_{\xi_{1}} = y_{i}-\mathtt{a_{2}}/4$ where 
$\mathtt{a_{2}} = -\left( a_{1}^{d} + a_{2}^{d} + a_{3}^{d} + a_{4}^{d}\right)$ and the values $y_{i}$ are given by $y_{1,2}=\frac{1}{2}\left(-k_{2}\pm \sqrt{k_{2}^2-4m_{2}}\right)$, 
$y_{3,4}=\frac{1}{2}\left(+k_{2}\pm \sqrt{k_{2}^2-4n_{2}}\right)$. Where $k_{2}$, $m_{2}$ and $n_{2}$ can be obtained from \ref{Biquadsoln} using 
$\mathtt{b_{2}} =  - 8 V^2 - 4 t^2 + a_{1}^{d} \left(a_{2}^{d} + a_{4}^{d}\right) + a_{3}^{d} \left(a_{1}^{d} + a_{4}^{d}\right) + a_{2}^{d} \left(a_{3}^{d} + a_{4}^{d}\right)$, 
$\mathtt{c_{2}} = 4 V^2 \left(a_{1}^{d} + a_{2}^{d} + a_{3}^{d} + a_{4}^{d}\right) - \left(a_{2}^{d} + a_{4}^{d}\right) \left(a_{1}^{d} a_{3}^{d} - 4 t^2\right) - a_{2}^{d} a_{4}^{d} \left(a_{1}^{d} + a_{3}^{d}\right)$ and  
$\mathtt{d_{2}} = 16 V^4 + 8 V^2 t \left(a_{2}^{d} - a_{4}^{d}\right) - 2 V^2 \left(a_{1}^{d} + a_{3}^{d}\right) \left(a_{2}^{d} + a_{4}^{d}\right) + a_{2}^{d} a_{4}^{d} \left(a_{1}^{d} a_{3}^{d} - 4 t^2\right)$. The corresponding normalized eigenvectors $\left|\lambda^{3,i}_{\xi_{1}}\right\rangle$ are given by 
\begin{eqnarray}
\left|\lambda^{3,i}_{\xi_{1}}\right\rangle=C^{1,i}_{\xi_{1}}\left|1\right\rangle_{3e}+C^{2,i}_{\xi_{1}}\left|3\right\rangle^{\prime}_{3e}+C^{3,i}_{\xi_{1}}\left|4\right\rangle^{\prime}_{3e}+C^{4,i}_{\xi_{1}}\left|7\right\rangle^{\prime}_{3e}. 
\label{gseta2}
\end{eqnarray}
The coefficients $C^{j,i}_{\xi_{1}}$ in a compact form are given as $C^{1,i}_{\xi_{1}}=\pm \left( \mathcal{K}_{3i}\mathcal{P}_{4i}\right)/\mathcal{D}_{1i}$, \break
$C^{2,i}_{\xi_{1}}=\mp \sqrt{2}V \mathcal{P}_{4i}\left(\mathcal{K}_{1i}-\mathcal{K}_{3i}\right)/\left(\mathcal{P}_{2i}\mathcal{D}_{1i}\right)$, 
$C^{3,i}_{\xi_{1}}=\pm \left(\mathcal{K}_{1i}\mathcal{P}_{4i} \right)/\mathcal{D}_{1i}$, 
$C^{4,i}_{\xi_{1}}=\pm \sqrt{2}V \left(\mathcal{K}_{1i}+\mathcal{K}_{3i}\right)/\mathcal{D}_{1i}$. Where 
$\mathcal{K}_{1i}\equiv\mathcal{P}_{2i}\left(a_{1}^{d}-2t-\lambda^{3,i}_{\xi_{1}}\right)-4V^{2}$, 
$\mathcal{P}_{2i}\equiv a_{2}^{d}-\lambda^{3,i}_{\xi_{1}}$, 
$\mathcal{K}_{3i}\equiv\mathcal{P}_{2i}\left(a_{3}^{d}-2t-\lambda^{3,i}_{\xi_{1}}\right)-4V^{2}$, 
$\mathcal{P}_{4i}\equiv a_{4}^{d}-\lambda^{3,i}_{\xi_{1}}$ with $\mathcal{D}_{1i}=\sqrt{\mathcal{P}_{4i}^{2}\left(\mathcal{K}_{1i}^{2}+\mathcal{K}_{3i}^{2}\right)+2V^{2}\left[\mathcal{P}_{4i}^{2}\left( a_{1}^{d}-a_{3}^{d}\right)^{2}+ \left(\mathcal{K}_{1i}+\mathcal{K}_{3i}\right)^{2}\right]}$.
\\
\indent
The other $4\times 4$ matrix defined over the basis $\left\lbrace \left|2\right\rangle_{3e}, \left|5\right\rangle^{\prime}_{3e}, \left|6\right\rangle^{\prime}_{3e}, \left|8\right\rangle^{\prime}_{3e} \right\rbrace$ is given as 
\begin{eqnarray*}
\left(
\begin{array}{cccc}
a_{1}^{d} & \sqrt{6}V & 0 & -\sqrt{6}V 
\\
\sqrt{6}V & a_{5}^{d} & \sqrt{2}V & 0 
\\
0 & \sqrt{2}V & a_{3}^{d} & -\sqrt{2}V 
\\
-\sqrt{6}V & 0 & -\sqrt{2}V & a_{6}^{d} 
\\
\end{array}
\right)
\end{eqnarray*}
where $a_{5}^{d}=3\varepsilon +t+U+2g$, 
$a_{6}^{d} = \varepsilon - t+2\varepsilon_{F}$. 
The eigenvalues of the above matrix labeled as $i=5,6,7,8$ are obtained as 
$\lambda^{3,i}_{\xi_{1}}=y_{i}-\mathtt{a_{3}}/4$ where 
$\mathtt{a_{3}} = - \left(a_{1}^{d} + a_{3}^{d} + a_{5}^{d} + a_{6}^{d}\right)$ and the values $y_{i}$ are given by $y_{5,6}=\frac{1}{2}\left(-k_{3}\pm \sqrt{k_{3}^2-4m_{3}}\right)$ and  
$y_{7,8}=\frac{1}{2}\left(+k_{3}\pm \sqrt{k_{3}^2-4n_{3}}\right)$. Where $k_{3}$, $m_{3}$ and $n_{3}$ can be obtained from \ref{Biquadsoln} using $\mathtt{b_{3}} = a_{1}^{d} a_{5}^{d} + a_{3}^{d} a_{6}^{d} + \left(a_{1}^{d} + a_{5}^{d}\right) \left(a_{3}^{d} + a_{6}^{d}\right)- 16 V^2$, $\mathtt{c_{3}} = 4 V^2 \left(a_{1}^{d} + 3 a_{3}^{d}\right) - \left(a_{5}^{d} + a_{6}^{d}\right) \left(a_{1}^{d} a_{3}^{d} - 8 V^2\right) - a_{5}^{d} a_{6}^{d} \left(a_{1}^{d} + a_{3}^{d}\right)$ and $\mathtt{d_{3}} = a_{1}^{d} a_{3}^{d} a_{5}^{d} a_{6}^{d} - 2 V^2 \left(a_{5}^{d} + a_{6}^{d}\right) \left(a_{1}^{d} + 3 a_{3}^{d}\right)$. 
The corresponding normalized eigenvectors $\left|\lambda^{3,i}_{\xi_{1}}\right\rangle$ are given as 
\begin{eqnarray}
\left|\lambda^{3,i}_{\xi_{1}}\right\rangle=C^{5,i}_{\xi_{1}}\left|2\right\rangle_{3e}+C^{6,i}_{\xi_{1}}\left|5\right\rangle^{\prime}_{3e}+C^{7,i}_{\xi_{1}}\left|6\right\rangle^{\prime}_{3e}+C^{8,i}_{\xi_{1}}\left|8\right\rangle^{\prime}_{3e}.
\label{gseta3}
\end{eqnarray}
The coefficients $C^{j,i}_{\xi_{1}}$ are given as
$C^{5,i}_{\xi_{1}}=\pm \sqrt{6}V\mathcal{P}_{3i}^{\prime}\left(\mathcal{P}_{5i}^{\prime}+\mathcal{P}_{6i}^{\prime}\right)/\mathcal{D}_{2i}$, 
$C^{6,i}_{\xi_{1}}=\mp \mathcal{P}_{1i}^{\prime}\mathcal{P}_{3i}^{\prime}\mathcal{P}_{6i}^{\prime}/\mathcal{D}_{2i}$, 
$C^{7,i}_{\xi_{1}}=\pm \sqrt{2}V\mathcal{P}_{1i}^{\prime}\left(\mathcal{P}_{5i}^{\prime}+\mathcal{P}_{6i}^{\prime} \right)/\mathcal{D}_{2i}$, 
$C^{8,i}_{\xi_{1}}=\pm \mathcal{P}_{1i}^{\prime}\mathcal{P}_{3i}^{\prime}\mathcal{P}_{5i}^{\prime}/\mathcal{D}_{2i}$, $\mathcal{P}_{1i}^{\prime}=a_{1}^{d}-\lambda^{3,i}_{\xi_{1}}$, 
$\mathcal{P}_{3i}^{\prime}=a_{3}^{d}-\lambda^{3,i}_{\xi_{1}}$, 
$\mathcal{P}_{5i}^{\prime}=a_{5}^{d}-\lambda^{3,i}_{\xi_{1}}$, 
$\mathcal{P}_{6i}^{\prime}=a_{6}^{d}-\lambda^{3,i}_{\xi_{1}}$ and 
$\mathcal{D}_{2i}=\sqrt{\mathcal{P}_{1i}^{\prime 2}\mathcal{P}_{3i}^{\prime 2}\left(\mathcal{P}_{5i}^{\prime 2}+\mathcal{P}_{6i}^{\prime 2}\right)
+2V^{2}\left(\mathcal{P}_{6i}^{\prime}+\mathcal{P}_{5i}^{\prime} \right)^{2}\left(3\mathcal{P}_{3i}^{\prime 2} +\mathcal{P}_{1i}^{\prime 2}\right)}$. 
The ground state of this subspace $\left|\lambda_{\xi_{1}}^{3,0}\right\rangle$ corresponds to $\lambda_{\xi_{1}}^{3,0} =\min\left(\lambda_{\xi_{1}}^{3,1}, \lambda_{\xi_{1}}^{3,5} \right)$. In this subspace two of the eigenvalues obtained in the limit $V\rightarrow 0$ with $\varepsilon_{F}=0$ as $\lambda^{1}_{dqd}=3\varepsilon-t+U+2g$ and $\lambda^{2}_{dqd}=3\varepsilon+t+U+2g$ correspond to the eigenvalues of the isolated DQDs.
\subsubsection{Infinite $U\rightarrow\infty$ limit}
The eigenstates calculated for the subspace $ \xi_{1}\equiv\left(N,S,S_{z}\right)=\left(3,\frac{1}{2},+\frac{1}{2}\right)$ in previous section \ref{h3site3eS1by2} do not consider any limiting case for any of the system parameters. 
In this section we consider the infinite ondot Coulomb interaction $U\rightarrow\infty$ limit to calculate the eigenstate of the Hamiltonian ${\bf H}_{3-site}$ for the same subspace. 
Considering $U\rightarrow\infty$ limit we see that the Hilbert space dimensionality reduces from eight to four as the basis states $\left|3\right\rangle_{3e}$, $\left|4\right\rangle_{3e}$, $\left|5\right\rangle_{3e}$ and $\left|6\right\rangle_{3e}$ having the double occupancy on the dots are eliminated. 
Thus we have to consider only four dimensional Hilbert space with basis states $\left|1\right\rangle_{3e}$, $\left|2\right\rangle_{3e}$, $\left|7\right\rangle_{3e}$ and $\left|8\right\rangle_{3e}$.
The Hamiltonian matrix over the reduced four dimensional Hilbert space becomes
\begin{eqnarray*}
\left(
\begin{array}{cccc}
a_{1}^{d} & 0 & -V & -V 
\\
0 & a_{1}^{d} & -\sqrt{3} V &\sqrt{3} V 
\\
-V & -\sqrt{3}V & \varepsilon_{4}^{\prime\prime} & t 
\\
-V & \sqrt{3}V & t & \varepsilon_{4}^{\prime\prime} 
\end{array}
\right).
\end{eqnarray*}
The above matrix under the basis transformation $\left|7\right\rangle_{3e}^{\prime}=\frac{1}{\sqrt{2}}\left(\left|7\right\rangle_{3e}+\left|8\right\rangle_{3e} \right)$ and \break $\left|8\right\rangle_{3e}^{\prime}=\frac{1}{\sqrt{2}}\left(\left|7\right\rangle_{3e}-\left|8\right\rangle_{3e} \right)$, block diagonalizes into two $2\times 2$ matrices of which one defined over the basis $\left\lbrace \left|1\right\rangle_{3e},\left|7\right\rangle_{3e}^{\prime} \right\rbrace$ is given by 
\begin{eqnarray*}
\left(
\begin{array}{cc}
a_{1}^{d} & -\sqrt{2}V
\\
-\sqrt{2}V & a_{4}^{d}
\end{array}
\right)
\end{eqnarray*}
The eigenvalues are given as 
$\lambda_{\xi_{1},\infty}^{3,j}=\frac{1}{2}\left(a_{1}^{d}+a_{4}^{d}+\alpha^{j}\sqrt{\left(a_{1}^{d}-a_{4}^{d}\right)^{2}+8V^{2}} \right)$ labeled by $j=1,2$  with $\alpha^{j}=-1$ for $j=1$ and $\alpha^{j}=+1$ for $j=2$. The corresponding eigenvectors are given as 
$\left|\lambda_{\xi_{1},\infty}^{3,j}\right\rangle=C_{\xi_{1},\infty}^{1,j}\left|2\right\rangle_{3e}+C_{\xi_{1},\infty}^{2,j}\left|8\right\rangle_{3e}^{\prime}$. The coefficients $C_{\xi_{1},\infty}^{i,j}$ are obtained as
$C_{\xi_{1},\infty}^{j,1}=\sqrt{2}V/\mathcal{D_{\infty}}$ and $C_{\xi_{1},\infty}^{j,1}=\left(a_{1}^{d}-\lambda_{\xi_{1},\infty}^{3,j}\right)/\mathcal{D_{\infty}}$ with $\mathcal{D_{\infty}}=\sqrt{\left(a_{1}^{d}-\lambda_{\xi_{1},\infty}^{3,j}\right)+2V^{2}}$.
\\
\indent
The other $2\times 2$ matrix defined over the basis $\left\lbrace \left|2\right\rangle_{3e},\left|8\right\rangle_{3e}^{\prime} \right\rbrace$ is given by 
\begin{eqnarray*}
\left(
\begin{array}{cc}
a_{1}^{d} & -\sqrt{6}V
\\
-\sqrt{6}V & a_{6}^{d}
\end{array}
\right)
\end{eqnarray*}
where $\varepsilon^{\prime\prime}_{4}$, $a_{1}^{d}$ and $a_{6}^{d}$ are defined in section \ref{h3site3eS1by2}. The eigenvalues $\lambda_{\xi_{1},\infty}^{3,k}$ labeled by $k=3,4$ are given as 
$\lambda_{\xi_{1},\infty}^{3,k}=\frac{1}{2}\left(a_{1}^{d}+a_{6}^{d}+\alpha^{k}\sqrt{\left(a_{1}^{d}-a_{6}^{d}\right)^{2}+24V^{2}} \right)$ with $\alpha^{k}=-1$ for $k=3$ and $\alpha^{k}=+1$ for $k=4$. 
The coefficients $C_{\xi_{1},\infty}^{i,j}$ are given as
$C_{\xi_{1},\infty}^{j,1}=\sqrt{6}V/\mathcal{D_{\infty}}$ and $C_{\xi_{1},\infty}^{j,1}=\left(a_{1}^{d}-\lambda_{\xi_{1},\infty}^{3,j}\right)/\mathcal{D_{\infty}}$ with $\mathcal{D_{\infty}}=\sqrt{\left(a_{1}^{d}-\lambda_{\xi_{1},\infty}^{3,j}\right)+6V^{2}}$ .\\
\indent
It can be seen that if the Fermi levels of the leads are set equal to zero {\it i.e.} $\varepsilon_{F}=0$ and ondot energies taken below it $\varepsilon<\varepsilon_{F}$, the ground state corresponds to $\left|\lambda_{\xi_{1},\infty}^{3,3}\right\rangle$ with eigenvalue $\lambda_{\xi_{1},\infty}^{3,3}$.
The four particle ground state of the complete zero bandwidth Hamiltonian ${\bf H}_{zbw}$ for $U\rightarrow \infty$ case can be found by adding an electron to the one particle ground state of the Hamiltonian ${\bf H}_{1-site}$ (given in section \ref{h3site1e}) and combining it with the ground state $\left|\lambda_{\xi_{1},\infty}^{3,3}\right\rangle$ to form a singlet with total spin $S=0$. Thus, the only ground state of the Hamiltonian ${\bf H}_{zbw}$ in the $U\rightarrow\infty$ limit is given by 
$\left|\lambda_{(4,0,0),\infty}^{4,0}=\lambda_{\xi_{1},\infty}^{3,3}+\varepsilon_{F}\right\rangle=\frac{1}{\sqrt{2}}\left(\alpha^{\dag}_{a\uparrow}\left|\lambda_{3,\frac{1}{2},-\frac{1}{2},\infty}^{3,3}\right\rangle-\alpha^{\dag}_{a\downarrow}\left|\lambda_{\xi_{1},\infty}^{3,3}\right\rangle \right)$. 
It is found that the spin-spin correlation between the dots in this ground state given by 
\begin{eqnarray} 
\left\langle \lambda_{(4,0,0),\infty}^{4,0} \right|{\bf S_{1}\cdot S_{2}}\left|\lambda_{(4,0,0),\infty}^{4,0}\right\rangle 
=\frac{1}{4}\left|C_{\xi_{1},\infty}^{1,3}\right|^{2} 
\label{corrUinf} 
\end{eqnarray} 
 is always ferromagnetic. 
\subsubsection{Subspace $N=3$, $S=\frac{3}{2}$, $S_{z}=\frac{3}{2}$}
\label{h3site3eS3by2}
The only eigenstate and corresponding eigenvector in this subspace is given as ${\bf H}_{3-site}\left|9\right\rangle_{3e} = \left(2\varepsilon+\varepsilon_{F}\right)\left|9\right\rangle_{3e}$ where 
$\left|9\right\rangle_{3e}=\alpha^{\dag}_{s\uparrow}\left|\sigma\right\rangle$.
\subsection{Four electron eigenstates of ${\bf H}_{3-site}$ Hamiltonian}
\label{h3site4e}
\subsubsection{Subspace $N=4$, $S=0$, $S_{z}=0$}
\label{h3site4eS0}
This subspace may be labeled by the triad $ \xi_{2} \equiv\left(N,S,S_{z}\right)=\left(4,0,0\right)$. The Hilbert space for the four-electron in the $S^{2}$-symmetry adapted basis with total spin $S=0$ is six dimensional. The basis states are given as $\left|1\right\rangle_{4e}= \alpha^{\dag}_{s\uparrow}\alpha^{\dag}_{s\downarrow}\left|\Xi\right\rangle$, 
$\left|2\right\rangle_{4e}= 
\frac{1}{\sqrt{2}}\left(c^{\dag}_{1\uparrow}\alpha^{\dag}_{s\downarrow}+\alpha^{\dag}_{s\uparrow}c^{\dag}_{1\downarrow}\right)
\left|\bar{i}\right\rangle$, 
$\left|3\right\rangle_{4e}= 
\frac{1}{\sqrt{2}}\left(c^{\dag}_{2\uparrow}\alpha^{\dag}_{s\downarrow}+\alpha^{\dag}_{s\uparrow}c^{\dag}_{2\downarrow}\right)
\left|i\right\rangle$, 
$\left|4\right\rangle_{4e}= \left|D\right\rangle$, 
$\left|5\right\rangle_{4e}= \alpha^{\dag}_{s\uparrow}\alpha^{\dag}_{s\downarrow}\left|i\right\rangle$, 
$\left|6\right\rangle_{4e}= \alpha^{\dag}_{s\uparrow}\alpha^{\dag}_{s\downarrow}\left|\bar{i}\right\rangle$.
Where $\left|\Xi\right\rangle$, $\left|\bar{i}\right\rangle$, $\left|D\right\rangle$ and $\left|i\right\rangle$ are defined in {\ref{shorthand}.
\begin{eqnarray*}
\left(
\begin{array}{cccccc}
\mathit{e}_{1} & -\sqrt{2}V & -\sqrt{2}V & 0 & \sqrt{2}t & \sqrt{2}t 
\\
-\sqrt{2}V & \mathit{e}_{2} & -t & 2V & 0 & 2V 
\\
-\sqrt{2}V & -t & \mathit{e}_{2} & 2V & 2V & 0 
\\
 0 & 2V & 2V & \mathit{e}_{3} & 0 & 0 
 \\
\sqrt{2}t & 0 & 2V & 0 & \mathit{e}_{4} & 0 
\\
\sqrt{2}t & 2V & 0 & 0 & 0 & \mathit{e}_{4}
\end{array}
\right)
\end{eqnarray*}
Where $\mathit{e}_{1}=2\varepsilon+g+2\varepsilon_{F}$, $\mathit{e}_{2}=3\varepsilon+U+2g+\varepsilon_{F}$,   $\mathit{e}_{3}=4\varepsilon+2U+4g$ and $\mathit{e}_{4}=2\varepsilon+U+2\varepsilon_{F}$. The above matrix under the basis transformation 
$\left|2\right\rangle^{\prime}_{4e}=\frac{1}{\sqrt{2}}\left(\left|2\right\rangle_{4e}+\left|3\right\rangle_{4e} \right)$, 
$\left|3\right\rangle^{\prime}_{4e}=\frac{1}{\sqrt{2}}\left(\left|2\right\rangle_{4e}-\left|3\right\rangle_{4e} \right)$, 
$\left|5\right\rangle^{\prime}_{4e}=\frac{1}{\sqrt{2}}\left(\left|5\right\rangle_{4e}+\left|6\right\rangle_{4e} \right)$
and $\left|6\right\rangle^{\prime}_{4e}=\frac{1}{\sqrt{2}}\left(\left|5\right\rangle_{4e}-\left|6\right\rangle_{4e}\right)$ block diagonalizes into two matrices, one $4\times 4$ and other $2\times2$. The $4\times4$ matrix defined over the basis $\left\lbrace \left|1\right\rangle_{4e}, \left|2\right\rangle^{\prime}_{4e}, \left|4\right\rangle_{4e}, \left|5\right\rangle^{\prime}_{4e} \right\rbrace$ is given as
\begin{eqnarray*}
\left(
\begin{array}{cccc}
\mathit{e}_{1} & -2V & 0 & 2t  
\\
-2V & \mathit{e}_{2}^{\prime} & 2\sqrt{2}V & 2V  
\\
0 & 2\sqrt{2}V & \mathit{e}_{3} & 0  
\\
2t & 2V & 0 & \mathit{e}_{4}
\end{array}
\right)
\end{eqnarray*}
with $\mathit{e}_{2}^{\prime}=\mathit{e}_{2}-t$. The eigenvalues of the above matrix labeled by $i=1,2,3,4$ are obtained as $\lambda^{3,i}_{\xi_{2}}=y_{i}-\mathtt{a_{4}}/4$ where $\mathtt{a_{4}} = - \left(\mathit{e}_{1} + \mathit{e}_{2}^{\prime} + \mathit{e}_{3} + \mathit{e}_{4}\right)$. The values $y_{i}$ are given by
$y_{1,2}=\frac{1}{2}\left(-k_{4}\pm \sqrt{k_{4}^2-4m_{4}}\right)$, 
$y_{3,4}=\frac{1}{2}\left(+k_{4}\pm \sqrt{k_{4}^2-4n_{4}}\right)$ where $k_{4}$, $m_{4}$ and $n_{4}$ can be obtained from \ref{Biquadsoln} using the parameters $\mathtt{b_{4}} = -4\left(t^{2} + 4 V^2\right)  + \mathit{e}_{1} \mathit{e}_{4} + \mathit{e}_{2}^{\prime} \mathit{e}_{3} + \left(\mathit{e}_{1} + \mathit{e}_{4}\right) \left(\mathit{e}_{2}^{\prime} + \mathit{e}_{3}\right)$, $\mathtt{c_{4}} = 8V^2 \left(\mathit{e}_{3} + 2 t\right) - \left(\mathit{e}_{2}^{\prime} + \mathit{e}_{3}\right) \left(\mathit{e}_{1} \mathit{e}_{4} - 4 t^2\right) - \left(\mathit{e}_{1} + \mathit{e}_{4}\right) \left(\mathit{e}_{2}^{\prime} \mathit{e}_{3} - 12 V^2\right)$ and $\mathtt{d_{4}} = \left(\mathit{e}_{2}^{\prime} \mathit{e}_{3} - 8 V^2\right) \left(\mathit{e}_{1} \mathit{e}_{4} - 4 t^2\right) - 4 V^2 \mathit{e}_{3} \left(\mathit{e}_{1} + \mathit{e}_{4} + 4 t\right)$. The eigenvectors $\left|\lambda^{3,i}_{\xi_{2}}\right\rangle$ corresponding to the eigenvalues 
$\lambda^{3,i}_{\xi_{2}}$ are given as 
\begin{eqnarray}
\left|\lambda^{3,i}_{\xi_{2}}\right\rangle=C^{1,i}_{\xi_{2}}\left|1\right\rangle_{4e}+C^{2,i}_{\xi_{2}}\left|2\right\rangle^{\prime}_{4e}+C^{3,i}_{\xi_{2}}\left|4\right\rangle_{4e}+C^{4,i}_{\xi_{2}}\left|5\right\rangle^{\prime}_{4e}. 
\label{gseta4}
\end{eqnarray}
The coefficients $C^{j,i}_{\xi_{2}}$ are given by
$C^{1,i}_{\xi_{2}}=\mp \left(2V\mathcal{P}_{4i}^{\prime\prime}\mathcal{K}_{3i}^{\prime}\right)/\mathcal{D}_{3i} $, 
$C^{2,i}_{\xi_{2}}=\mp \left(\mathcal{K}_{3i}^{\prime}\mathcal{K}_{2i}^{\prime} \right)/\mathcal{D}_{3i}$, 
$C^{3,i}_{\xi_{2}}= \pm \left(2\sqrt{2}V\mathcal{K}_{2i}^{\prime} \right)/ \mathcal{D}_{3i}$, 
$C^{4,i}_{\xi_{2}}= \pm \left(2V\mathcal{P}_{1i}^{\prime\prime}\mathcal{K}_{3i}^{\prime}\right)/\mathcal{D}_{3i} $ where 
$\mathcal{K}_{1i}^{\prime}=\mathit{e}_{1}-\lambda^{3,i}_{\xi_{2}}$, 
$\mathcal{K}_{3i}^{\prime}=\mathit{e}_{3}-\lambda^{3,i}_{\xi_{2}}$, 
$\mathcal{K}_{4i}^{\prime}=\mathit{e}_{4}-\lambda^{3,i}_{\xi_{2}}$, 
$\mathcal{K}_{2i}^{\prime}=\mathcal{K}_{1i}^{\prime}\mathcal{K}_{4i}^{\prime}-4t^{2}$, 
$\mathcal{P}_{1i}^{\prime\prime}=\mathit{e}_{1}+2t-\lambda^{3,i}_{\xi_{2}}$, 
$\mathcal{P}_{4i}^{\prime\prime}=\mathit{e}_{4}+2t-\lambda^{3,i}_{\xi_{2}}$ and \break
$\mathcal{D}_{3i}=\sqrt{\left(\mathcal{K}_{3i}^{\prime 2}+8V^{2}\right)\mathcal{K}_{2i}^{\prime 2}+4V^{2}\mathcal{K}_{3i}^{\prime 2}
\left(\mathcal{P}_{1i}^{\prime\prime 2}+\mathcal{P}_{4i}^{\prime\prime 2} \right) }$. 
The other $2\times 2$ matrix defined over the basis 
$\left\lbrace \left|3\right\rangle^{\prime}_{4e}, \left|6\right\rangle^{\prime}_{4e} \right\rbrace$ is given as
\begin{eqnarray*}
\left(
\begin{array}{cc}
\mathit{e}_{2}^{\prime\prime} & -2V 
\\
-2V & \mathit{e}_{4} 
\\
\end{array}
\right).
\end{eqnarray*}
The eigenvalues are given as 
$\lambda_{\xi_{2}}^{3,5}=\frac{1}{2}\left(\mathit{e}_{4}+\mathit{e}_{2}^{\prime\prime}-R_{1} \right)$ and 
$\lambda_{\xi_{2}}^{3,6}=\frac{1}{2}\left(\mathit{e}_{4}+\mathit{e}_{2}^{\prime\prime}+R_{1} \right)$ where  $\mathit{e}_{2}^{\prime\prime}=\mathit{e}_{2}+t$ and $R_{1}=\sqrt{\left(\mathit{e}_{4}-\mathit{e}_{2}^{\prime\prime} \right)^{2}+16V^{2}}$. 
The corresponding eigenvectors labeled by $i=5,6$ are given as 
$\left|\lambda_{\xi_{2}}^{3,i}\right\rangle = \pm \frac{1}{\sqrt{\left(\mathit{e}_{2}^{\prime\prime}-\lambda_{\xi_{2}}^{3,i} \right)^{2}+4V^{2}}}\left[ 2V\left|3\right\rangle^{\prime}_{4e} 
+ \left(\mathit{e}_{2}^{\prime\prime}-\lambda_{\xi_{2}}^{3,i} \right)\left|6\right\rangle^{\prime}_{4e} \right]$. In the limit $V\rightarrow 0$ with $\varepsilon_{F}=0$ one of the above eigenvalues correspond to the isolated DQDs as $\lambda_{dqd}^{1}=4\varepsilon+2U+4g$.
\subsubsection{Subspace $N=4$, $S=1$, $S_{z}=1$}
\label{h3site4eS1}
This subspace may be labeled by $\xi_{2}^{\prime}=\left( N,S,S_{z} \right)=\left( 4,1,1\right)$. The Hilbert space is three dimensional. The basis states are given as 
$\left|7\right\rangle_{4e}=c^{\dag}_{2\uparrow}\alpha^{\dag}_{s\uparrow}\left|i\right\rangle$, 
$\left|8\right\rangle_{4e}=c^{\dag}_{1\uparrow}\alpha^{\dag}_{s\uparrow}\left|\bar{i}\right\rangle$ and 
$\left|9\right\rangle_{4e}=\alpha^{\dag}_{s\uparrow}\alpha^{\dag}_{s\downarrow}\left|\sigma\right\rangle$.
The Hamiltonian matrix over the above three dimensional Hilbert space becomes
\begin{eqnarray*}
\left(
\begin{array}{ccc}
\mathit{e}_{1}^{\prime} & -t & \sqrt{2}V 
\\
-t & \mathit{e}_{1}^{\prime} & -\sqrt{2}V 
\\
\sqrt{2}V & -\sqrt{2}V & \mathit{e}_{3}^{\prime}
\end{array}
\right)
\end{eqnarray*}
where $\mathit{e}_{1}^{\prime}=3\varepsilon+U+2g+\varepsilon_{F}$, $\mathit{e}_{3}^{\prime}=2\varepsilon+g+2\varepsilon_{F}$.
The above matrix can be easily diagonalized by performing the transformation $\left|7\right\rangle^{\prime}_{4e}=\frac{1}{\sqrt{2}}\left(\left|7\right\rangle_{4e}+\left|8\right\rangle_{4e}\right)$ and $\left|8\right\rangle^{\prime}_{4e}=\frac{1}{\sqrt{2}}\left(\left|7\right\rangle_{4e}-\left|8\right\rangle_{4e}\right)$. The eigenvalues are given as 
$\lambda_{\xi_{2}^{\prime}}^{3,7}=\mathit{e}_{1}^{\prime}-t$, 
$\lambda_{\xi_{2}^{\prime}}^{3,8}=\frac{1}{2}\left(\mathit{e}_{3}^{\prime}+\mathit{e}_{1}^{\prime}+t-R_{2} \right)$ and $\lambda_{\xi_{2}^{\prime}}^{3,9}=\frac{1}{2}\left(\mathit{e}_{3}^{\prime}+\mathit{e}_{1}^{\prime}+t+R_{2} \right)$. 
The corresponding eigenvectors are given by
 $\left|\lambda_{\xi_{2}^{\prime}}^{3,7}\right\rangle=\left|7\right\rangle^{\prime}_{4e}$ 
 and the other two labeled by $k=8,9$ as 
 $\left|\lambda_{\xi_{2}^{\prime}}^{3,k}\right\rangle=\frac{\pm 1}{\sqrt{\left(\lambda_{\xi_{2}^{\prime}}^{3,k}-\mathit{e}_{3}^{\prime}\right)^{2}+4V^{2}}}\left[ \left( \lambda_{\xi_{2}^{\prime}}^{3,k}-\mathit{e}_{3}^{\prime}\right)\left|8\right\rangle^{\prime}_{4e}+2V\left|9\right\rangle_{4e} \right]$ 
with $R_{2}=\sqrt{\left(\mathit{e}_{1}^{\prime}+t-\mathit{e}_{3}^{\prime}\right)^{2}+16V^{2}}$. The ground state in this subspace is given by $\left|\lambda_{\xi_{2}^{\prime}}^{3,8}\right\rangle$.
\subsection{Five electron eigenstates of ${\bf H}_{3-site}$ Hamiltonian}
\label{h3site5e}
With five electron ($N=5$) on three site, the only possible total spin is $S=1/2$. The subspace may be labeled as 
$\xi_{3}=\left( N,S,S_{z}\right)=\left( 5,1/2,+1/2\right)$. The Hilbert space is three dimensional. The basis states are given as 
$\left|1\right\rangle_{5e} = \alpha^{\dag}_{s\uparrow}\left|D\right\rangle$, 
$\left|2\right\rangle_{5e} = c^{\dag}_{2\uparrow}\alpha^{\dag}_{s\uparrow}\alpha^{\dag}_{s\downarrow}\left|i\right\rangle$ 
and 
$\left|3\right\rangle_{5e} = c^{\dag}_{1\uparrow}\alpha^{\dag}_{s\uparrow}\alpha^{\dag}_{s\downarrow}\left|\bar{i}\right\rangle$. The Hamiltonian matrix over the Hilbert space is given as
\begin{eqnarray*}
\left(
\begin{array}{ccc}
\mathit{e}_{1}^{d} & -\sqrt{2}V & -\sqrt{2}V 
\\
-\sqrt{2}V & \mathit{e}_{2}^{d} & -t 
\\
-\sqrt{2}V & -t & \mathit{e}_{2}^{d}
\end{array}
\right)
\end{eqnarray*}
where $\mathit{e}_{1}^{d}=4\varepsilon+2U+4g+\varepsilon_{F}$ and 
$\mathit{e}_{2}^{d}=3\varepsilon+U+2g+2\varepsilon_{F}$. The eigenvalues are given as $\lambda_{\xi_{3}}^{3,1}= \mathit{e}_{2}^{d}+t$,
$\lambda_{\xi_{3}}^{3,2}=\frac{1}{2}\left(\mathit{e}_{1}^{d}+\mathit{e}_{3}^{d}-R_{3} \right)$ and 
$\lambda_{\xi_{3}}^{3,3}=\frac{1}{2}\left(\mathit{e}_{1}^{d}+\mathit{e}_{3}^{d}+R_{3} \right)$. 
The eigenvectors are given as $\left|\lambda_{\xi_{3}}^{3,1}\right\rangle=\frac{1}{\sqrt{2}}\left(\left|2\right\rangle_{5e}-\left|3\right\rangle_{5e} \right)$ and 
$\left|\lambda_{\xi_{3}}^{3,k}\right\rangle = \pm \frac{1}{\sqrt{\left(\mathit{e}_{1}^{d}-\lambda_{\xi_{3}}^{3,k} \right)^{2}+4V^{2}}}\left[
 2V\left|1\right\rangle_{5e} + \left(\mathit{e}_{1}^{d}-\lambda_{\xi_{3}}^{3,k} \right)\left|2\right\rangle^{\prime}_{5e} \right]$ 
where $k=2,3$ with $\left|2\right\rangle^{\prime}_{5e} = \frac{1}{\sqrt{2}}\left(\left|2\right\rangle_{5e}+\left|3\right\rangle_{5e} \right)$, 
$\mathit{e}_{3}^{d}=\mathit{e}_{2}^{d}-t$ and $R_{3}=\sqrt{\left(\mathit{e}_{1}^{d}-\mathit{e}_{3}^{d} \right)^{2}+16V^{2}}$. 
The ground state in this subspace is given by $\left|\lambda_{\xi_{3}}^{3,2}\right\rangle$.
\subsection{Six electron eigenstate of ${\bf H}_{3-site}$ Hamiltonian}
\label{h3site6e}
With six electron in the system, the only possible total spin is $S=0$ and there is only one basis state which is also the eigenstate of the Hamiltonian ${\bf H}_{3-site}$ given as ${\bf H}_{3-site}\left|1\right\rangle_{6e}=(4\varepsilon+2U+4g+2\varepsilon_{F})\left|1\right\rangle_{6e}$ where 
$\left|1\right\rangle_{6e} = \alpha^{\dag}_{s\uparrow}\alpha^{\dag}_{s\downarrow}\left|D\right\rangle$.
\section{Spin-spin correlation for the half-filled case}
\label{analtcspincorr}
Using the eigenstates of the Hamiltonians ${\bf H}_{3-site}$ and ${\bf H}_{1-site}$ obtained analytically above, we now calculate spin-spin correlation $\left\langle{\bf S_{1}\cdot S_{2}}\right\rangle$ between the quantum dots for the half-filled case {\it i.e.} $N=4$ where ${\bf S_{1}}$ and ${\bf S_{2}}$ are the spins associated with dot-1 and dot-2, respectively. For the non-magnetic case, the ground state $\left|\lambda_{N=4,S=0,S_{z}=0}^{4,0}\right\rangle$ lies in total spin $S=0$ subspace. 
The Hilbert space dimensionality of the four-site problem corresponding to the zero-bandwidth Hamiltonian ${\bf H}_{zbw}$ in eq. (\ref{zerobandhamilp}) for the half-filled case in the subspace with total spin $S=0$ is $20$. Within this $20$ dimensional Hilbert space, the ground state lies in a four-dimensional subspace, identified as follows. 
The four possible ground states of the zero-bandwidth Hamiltonian ${\bf H}_{zbw}$ in eq. (\ref{zerobandhamilp}) can be constructed from the eigenstates of the Hamiltonians ${\bf H}_{3-site}$ in eq. ({\ref{3sitehp}}) and ${\bf H}_{1-site}$ in eq. ({\ref{1sitep}}) as below. 
\begin{enumerate}
\item  One way is to add two-electrons to the decoupled orbital ${\alpha_{a\sigma}}$ of the Hamiltonian ${\bf H}_{1-site}$ to form the singlet $\left|\lambda^{1,0}_{2,0,0}\right\rangle=\alpha_{a \uparrow}^{\dagger}\alpha_{a\downarrow}^{\dagger}\left|0\right\rangle$ and couple this to the two-electron singlet ground state $\left|\lambda_{2,0,0 }^{3,3}\right\rangle$ of the Hamiltonian ${\bf H}_{3-site}$ so as to form the $4$-electron singlet with total spin $S=0$.
 \item  Another way is to add one-electron to the decoupled orbital ${\alpha_{a\sigma}}$ of the Hamiltonian ${\bf H}_{1-site}$ to form the spin $\frac{1}{2}$ state  
 $\left|\lambda^{1,0}_{1,\frac{1}{2},+\frac{1}{2}}\right\rangle=\alpha_{a \uparrow}^{\dag}\left|0\right\rangle$ and couple this to the three-electron spin $\frac{1}{2}$ ground state $\left|\lambda_{3,\frac{1}{2},+\frac{1}{2}}^{3,0}\right\rangle$ of the three-site Hamiltonian ${\bf H}_{3-site}$ so as to form the $4$-electron singlet with total spin $S=0$.
\item  The last possibility to construct the four-electron singlet ground state $\left|\lambda_{4,0,0}^{4,0}\right\rangle$ of the Hamiltonian ${\bf H}_{zbw}$ is given by the four-electron ground state of the three-site Hamiltonian ${\bf H}_{3-site}$ {\it i.e.} when there is no electron in decoupled orbital ${\alpha_{a\sigma}}$ of the Hamiltonian ${\bf H}_{1-site}$.
\end{enumerate}
\begin{table}[!htb]
\resizebox{\linewidth}{!}{
\begin{tabular}{ccclll}
\hline \hline
${\eta}$ & $N^{1-site}$ & $N^{3-site}$ & Ground state & Ground state  & Possible ground states \\
         & & & of ${\bf H}_{1-site}$ & of ${\bf H}_{3-site}$ & $\left|\lambda^{4,\eta}_{\xi_{2}}\right\rangle$ of ${\bf H}_{zbw}$ \\
\hline \hline
 1 & 2 & 2 & $\left|\lambda_{\xi}^{1,0}=2\varepsilon_{F}\right\rangle$ & $\left|\lambda_{\xi}^{3,3}\right\rangle$ &  $\left|\lambda_{\xi_{2}}^{4,1} =\lambda_{\xi}^{3,3}+2\varepsilon_{F}\right\rangle=\alpha^{\dag}_{a\uparrow}\alpha^{\dag}_{a\downarrow}\left|\lambda_{\xi}^{3,3}\right\rangle$\\
 \medskip
 2 & 1 & 3 & $\left|\lambda_{\xi_{0}}^{1,0}=\varepsilon_{F}\right\rangle$ &$\left|\lambda_{\xi_{1}}^{3,1}\right\rangle$ &	$\left|\lambda_{\xi_{2}}^{4,2}=\lambda_{\xi{1}}^{3,1}+\varepsilon_{F}\right\rangle=\frac{1}{\sqrt{2}}\left(\alpha^{\dag}_{a\uparrow}\left|\lambda_{3,1/2,-1/2}^{3,1}\right\rangle-\alpha^{\dag}_{a\downarrow}\left|\lambda_{\xi{1}}^{3,1}\right\rangle \right)$\\
 \medskip
 3 & 1 & 3 & $\left|\lambda_{\xi_{0}}^{1,0}=\varepsilon_{F}\right\rangle$ & $\left|\lambda_{\xi_{1}}^{3,5}\right\rangle$ & $\left|\lambda_{\xi_{2}}^{4,3}=\lambda_{\xi{1}}^{3,5}+\varepsilon_{F}\right\rangle=\frac{1}{\sqrt{2}}\left(\alpha^{\dag}_{a\uparrow}\left|\lambda_{3,1/2,-1/2}^{3,5}\right\rangle-\alpha^{\dag}_{a\downarrow}\left|\lambda_{\xi{1}}^{3,5}\right\rangle \right)$\\
 \medskip
 4 & 0 & 4 & & $\left|\lambda_{\xi_{2}}^{3,1}\right\rangle$ & $\left|\lambda_{\xi_{2}}^{4,4}=\lambda_{\xi_{2}}^{3,1}\right\rangle=\left|\lambda_{\xi_{2}}^{3,1}\right\rangle$	\\
\hline \hline
 \end{tabular}}
\caption{Four possible ground states $\left|\lambda_{\xi_{2}}^{4,\eta}\right\rangle$ labeled by $\eta $ with corresponding eigenenergies $\lambda_{\xi_{2}}^{4,\eta}$ of the zero-bandwidth Hamiltonian ${\bf H}_{zbw}$ for four-electron half-filled case constructed using the ground states of the Hamiltonians ${\bf H}_{1-site}$ and ${\bf H}_{3-site}$. The eigenstates 
$\left|\lambda^{1,0}_{\xi_{0}}\right\rangle$ and $\left|\lambda^{1,0}_{\xi}\right\rangle$, with respective eigenvalues $\lambda^{1,0}_{\xi_{0}}$ and $\lambda^{1,0}_{\xi}$, correspond to one-electron and two-electron ground states of the Hamiltonian ${\bf H}_{1-site}$ given in section(\ref{h3site1e}). 
The eigenstates $\left|\lambda^{3,3}_{\xi}\right\rangle$, 
$\left(\left|\lambda^{3,1}_{\xi_{1}}\right\rangle, \left|\lambda^{3,5}_{\xi_{1}}\right\rangle\right)$ 
and $\left|\lambda^{3,1}_{\xi_{2}}\right\rangle$ are the two-electron, three-electron and four-electron ground states of the Hamiltonian ${\bf H}_{3-site}$ given in sections (\ref{h3site2eS0}), (\ref{h3site3eS1by2}) and (\ref{h3site4eS0}), respectively where $\lambda^{3,3}_{\xi}$, $\left( \lambda^{3,1}_{\xi_{1}}, \lambda^{3,5}_{\xi_{1}}\right)$ and $\lambda^{3,1}_{\xi_{2}}$ are the respective eigenvalues. The total number of electrons is given by $N=N^{1-site}+N^{3-site}$ where $N^{1-site}$ and $N^{3-site}$ are the number of electrons for the Hamiltonians ${\bf H}_{1-site}$ and ${\bf H}_{3-site}$, respectively. The triads $\xi_{0}= \left(1,\frac{1}{2},+\frac{1}{2}\right)$, $\xi = \left( 2,0,0\right)$, $\xi_{1}=\left(3,\frac{1}{2},+\frac{1}{2}\right)$ and $\xi_{2}=\left(4,0,0\right)$ are the set of quantum numbers $\left(N,S,S_{z}\right)$ labeling the eigenstates in $N$ particle sector; $S$ and $S_{z}$ being the total spin and its $z$-component, respectively.
\hfill \break}
\medskip
\label{tab:gsfe}  
\end{table}
\indent
 Table \ref{tab:gsfe} summarizes how the four possible ground states of the four-site Hamiltonian ${\bf H}_{zbw}$ at the half-filling can be constructed from the eigenstates of ${\bf H}_{3-site}$ and ${\bf H}_{1-site}$. For a given set of values of system parameters, the ground state $\left|\lambda_{\xi_{2}}^{4,\eta }\right\rangle$ of the four-electron system corresponds to the eigenvalue $\lambda_{\xi_{2}}^{4,\eta }=\min\left[ \lambda_{\xi_{2}}^{4,1}, \left(\lambda_{\xi_{2}}^{4,2}, \lambda_{\xi_{2}}^{4,3}\right),\lambda_{\xi_{2}}^{4,4}\right]$. One can find $\left|\lambda_{3,\frac{1}{2},-\frac{1}{2}}^{3,1}\right\rangle$ or $\left|\lambda_{3,\frac{1}{2},-\frac{1}{2}}^{3,5}\right\rangle$ using total spin lowering operator as $\left|\lambda_{3,\frac{1}{2},-\frac{1}{2}}^{3,5}\right\rangle=S^{-}\left|\lambda_{3,\frac{1}{2},+\frac{1}{2}}^{3,5}\right\rangle$ where $S^{-}=S_{1}^{-}+S_{2}^{-}+S_{s}^{-}$  with $S^{-}_{s}=\alpha^{\dag}_{s\downarrow}\alpha_{s\uparrow}$, for the symmetric combination of the leads. The form of these possible ground states are given as
\begin{small}
\begin{eqnarray}
\left|\lambda_{\xi_{2}}^{4,1} \right\rangle &=&
C^{1,3}_{\xi}\alpha^{\dag}_{a\uparrow}\alpha^{\dag}_{a\downarrow}\left|\Xi\right\rangle
+C^{2,3}_{\xi}\left(\frac{1}{\sqrt{2}}\right)\alpha^{\dag}_{a\uparrow}\alpha^{\dag}_{a\downarrow}\left(\left|i\right\rangle+\left|\bar{i}\right\rangle \right)
\nonumber \\
&+&
C^{3,3}_{\xi}\left(\frac{1}{2}\right)\left[
\left(c^{\dag}_{1\uparrow}+c^{\dag}_{2\uparrow} \right)\alpha^{\dag}_{s\downarrow}+\alpha^{\dag}_{s\uparrow}\left(c^{\dag}_{1\downarrow}+c^{\dag}_{2\downarrow} \right)
\right]\left|\bar{l}\right\rangle
+
C^{4,3}_{\xi}\alpha^{\dag}_{a\uparrow}\alpha^{\dag}_{a\downarrow}\left|l\right\rangle
\label{gs1}
\end{eqnarray}
\begin{eqnarray}
\left|\lambda_{\xi_{2}}^{4,2} \right\rangle &=&
C^{1,1}_{\xi_{1}}\left(\frac{1}{\sqrt{2}}\right)\left(\alpha^{\dag}_{a\uparrow}\alpha^{\dag}_{s\downarrow}-\alpha^{\dag}_{a\downarrow}\alpha^{\dag}_{s\uparrow} \right)\left|\Xi \right\rangle
+
C^{3,1}_{\xi_{1}}\left( \frac{1}{2}\right)\left(\alpha^{\dag}_{a\uparrow}\alpha^{\dag}_{s\downarrow}-\alpha^{\dag}_{a\downarrow}\alpha^{\dag}_{s\uparrow} \right)\left(\left|i\right\rangle+\left|\bar{i}\right\rangle\right)
\nonumber \\
&+&
C^{2,1}_{\xi_{1}}\left( \frac{1}{2}\right)\left[\left(\alpha^{\dag}_{a\uparrow}c^{\dag}_{2\downarrow}-\alpha^{\dag}_{a\downarrow}c^{\dag}_{2\uparrow}\right)\left|i\right\rangle
+\left(\alpha^{\dag}_{a\uparrow}c^{\dag}_{1\downarrow}-\alpha^{\dag}_{a\downarrow}c^{\dag}_{1\uparrow}\right)\left|\bar{i}\right\rangle\right]
\nonumber \\
&+&
C^{4,1}_{\xi_{1}}\left( \frac{1}{2}\right)\left[\alpha^{\dag}_{a\uparrow}\left(c^{\dag}_{1\downarrow}+c^{\dag}_{2\downarrow}\right)-\alpha^{\dag}_{a\downarrow}\left( c^{\dag}_{1\uparrow}+c^{\dag}_{2\uparrow}\right)
\right]\left|l\right\rangle 
\label{gs2}
\end{eqnarray} 
\begin{eqnarray}
\left|\lambda_{\xi_{2}}^{4,3} \right\rangle &=&
C^{5,5}_{\xi_{1}}\left(\frac{1}{\sqrt{6}}\right)\left[
\left(\alpha^{\dag}_{s\uparrow}\alpha^{\dag}_{a\downarrow}+\alpha^{\dag}_{s\downarrow}\alpha^{\dag}_{a\uparrow}\right) \left|\Theta \right\rangle+\sqrt{2}\alpha^{\dag}_{a\uparrow}\alpha^{\dag}_{s\uparrow}\left|\bar{\sigma}\right\rangle+\sqrt{2}\alpha^{\dag}_{a\downarrow}\alpha^{\dag}_{s\downarrow}\left|\sigma\right\rangle 
\right] 
\nonumber \\
&+&
C^{6,5}_{\xi_{1}}\left(\frac{1}{2}\right)\left[
\left(\alpha^{\dag}_{a\uparrow}c^{\dag}_{2\downarrow}-\alpha^{\dag}_{a\downarrow}c^{\dag}_{2\uparrow}\right)\left|i\right\rangle+
\left(\alpha^{\dag}_{a\downarrow}c^{\dag}_{1\uparrow}-\alpha^{\dag}_{a\uparrow}c^{\dag}_{1\downarrow} \right)\left|\bar{i}\right\rangle
\right] \nonumber \\
&+&
C^{7,5}_{\xi_{1}}\left(\frac{1}{2}\right)
\left(\alpha^{\dag}_{a\uparrow}\alpha^{\dag}_{s\downarrow}-\alpha^{\dag}_{a\downarrow}\alpha^{\dag}_{s\uparrow} \right)\left(\left|i\right\rangle- \left|\bar{i}\right\rangle\right)  +
\nonumber \\
&+&
C^{8,5}_{\xi_{1}}\left(\frac{1}{2}\right)\left[
\alpha^{\dag}_{a\uparrow}\left(c^{\dag}_{1\downarrow}-c^{\dag}_{2\downarrow} \right)-\alpha^{\dag}_{a\downarrow}\left(c^{\dag}_{1\uparrow}-c^{\dag}_{2\uparrow}\right)
\right]\left|l\right\rangle \nonumber \\
\label{gs3}
\end{eqnarray} 
\begin{eqnarray}
\left|\lambda_{\xi_{2}}^{4,4} \right\rangle &=&
C^{1,1}_{\xi_{2}}\alpha^{\dag}_{s\uparrow}\alpha^{\dag}_{s\downarrow}\left|\Xi\right\rangle 
+ C^{2,1}_{\xi_{2}}\left(\frac{1}{2}\right)\left[\left(c^{\dag}_{1\uparrow}\alpha^{\dag}_{s\downarrow}+\alpha^{\dag}_{s\uparrow}c^{\dag}_{1\downarrow} \right)\left|\bar{i}\right\rangle+\left(c^{\dag}_{2\uparrow}\alpha^{\dag}_{s\downarrow}+\alpha^{\dag}_{s\uparrow}c^{\dag}_{2\downarrow}\right)\left|i\right\rangle
\right]
\nonumber \\
&+&
C^{3,1}_{\xi_{2}}\left|D\right\rangle
+
C^{4,1}_{\xi_{2}}\left(\frac{1}{\sqrt{2}}\right) \alpha^{\dag}_{s\uparrow}\alpha^{\dag}_{s\downarrow} \left(\left|i\right\rangle +\left|\bar{i}\right\rangle \right).
\label{gs4}
\end{eqnarray}
\end{small}
Where the symbols $\left|i\left(\bar{i}\right)\right\rangle$, $\left|\sigma\left(\bar{\sigma}\right)\right\rangle$, $\left|\Theta\right\rangle$, $\left|\Xi\right\rangle$ and $\left|l\left(\bar{l}\right)\right\rangle$ represents states on the dots and the leads respectively, has been defined in \ref{shorthand}. At zero temperature, the spin-spin correlation between the dots corresponding to four possible ground states is calculated as $\left\langle\lambda^{4,\eta }_{\xi_{2}}\right|{\bf S_{1}.S_{2}}\left|\lambda^{4,\eta }_{\xi_{2}}\right\rangle$ where ${\bf S_{i}.S_{j}}=\frac{1}{2}\left(S_{i}^{+}S_{j}^{-}+S_{i}^{-}S_{j}^{+}\right)+S_{i}^{z}S_{j}^{z} $ with $S_{i}^{z}=\frac{1}{2}\left(n_{i\uparrow}-n_{i\downarrow} \right)$, $S_{i}^{+}=c^{\dag}_{i\uparrow}c_{i\downarrow}$ and $S_{i}^{-}=c^{\dag}_{i\downarrow}c_{i\uparrow}$. The analytically calculated spin-spin correlation between the dots and the occupancies in the four possible ground states for the four-site half-filled case is summarized in Table \ref{tab:ksfe}. If $\left|\lambda_{\xi_{2}}^{4,1}\right\rangle$ is the ground state of the system, the dots have average occupancies $\left<n_{i}\right>$ varying between $0$ to $1$, leading to antiferromagnetic correlation between the dots. The ground state $\left|\lambda_{\xi_{2}}^{4,2}\right\rangle$ also lead to antiferromagnetic correlation between the dots. The only ground state $\left|\lambda_{\xi_{2}}^{4,3}\right\rangle$ lead to ferromagnetic correlation between the dots and the average occupancies of the dots $\left<n_{i}\right>$ varies between $0$ and $1.5$. The average occupancies $\left<n_{i\sigma}\right>$ for the dots in the ground state $\left|\lambda_{\xi_{2}}^{4,4 }\right\rangle$ can have a maximum value 2 due to the doublet $\left|D\right>$ in eq. ({\ref{gs4}}) and the spin-spin correlation $\left<{\bf S_{1}\cdot S_{2}}\right>$ leading to antiferromagnetic correlation between dots to a maximum value of $-\frac{3}{4}$.   
\begin{table}[!htb]
\centering
\begin{tabular}{lcrl}
\hline \hline
${\eta}$& $\left\langle\lambda^{4,\eta}_{\xi_{2}}\right|{\bf S_{1}.S_{2}}\left|\lambda^{4,\eta}_{\xi_{2}}\right\rangle$  & Type of correlation & $\left\langle\lambda^{4,\eta}_{\xi_{2}}\right|n_{1\sigma}\left|\lambda^{4,\eta}_{\xi_{2}}\right\rangle$ \\ \hline \hline
 1 & $-\frac{3}{4}\left|C^{1,3}_{\xi}\right|^{2} $ & Antiferromagnetic & $\left|C^{1,3}_{\xi}\right|^{2}+\left|C^{2,3}_{\xi}\right|^{2}+\frac{1}{2}\left|C^{3,3}_{\xi}\right|^{2}$ \\
 2 &$-\frac{3}{4}\left|C^{1,1}_{\xi_{1}}\right|^{2} $ & Antiferromagnetic & $\left|C^{1,1}_{\xi_{1}}\right|^{2}+\frac{3}{2}\left|C^{2,1}_{\xi_{1}}\right|^{2}+\left|C^{3,1}_{\xi_{1}}\right|^{2}+\frac{1}{2}\left|C^{4,1}_{\xi_{1}}\right|^{2}$  \\
 3 &$+\frac{1}{4}\left|C^{5,5}_{\xi_{1}}\right|^{2} $ & Ferromagnetic	& $\left|C^{5,5}_{\xi_{1}}\right|^{2}+\frac{3}{2}\left|C^{6,5}_{\xi_{1}}\right|^{2}+\left|C^{7,5}_{\xi_{1}}\right|^{2}+\frac{1}{2}\left|C^{8,5}_{\xi_{1}}\right|^{2}$ \\
 4 &$-\frac{3}{4}\left|C^{1,1}_{\xi_{2}}\right|^{2} $ & Antiferromagnetic & $\left|C^{1,1}_{\xi_{2}}\right|^{2}+\frac{3}{2}\left|C^{2,1}_{\xi_{2}}\right|^{2}+2\left|C^{3,1}_{\xi_{2}}\right|^{2}+\left|C^{4,1}_{\xi_{2}}\right|^{2}$	\\
\hline \hline
\end{tabular}
\caption{Spin-spin correlation $\left\langle {\bf S_{1}}\cdot {\bf S_{2}}\right\rangle$ between the dots for the four possible ground states $\left|\lambda_{\xi_{2}}^{4,\eta} \right\rangle$ labeled by $\eta=1,2,3,4$ listed in Table {\ref{tab:gsfe}}. The coefficients given in the rows corresponding to $\eta=1,2,3,4$ {\it i.e.} $C^{1,3}_{\xi}$, $C^{1,1}_{\xi_{1}}$, $ C^{5,5}_{\xi_{1}}$ and $C^{1,1}_{\xi_{2}}$ etc. are given in equations (\ref{gseta1}), (\ref{gseta2}), (\ref{gseta3}) and  (\ref{gseta4}) respectively. The symbols $\xi_{0}= \left(1,\frac{1}{2},+\frac{1}{2}\right)$, $\xi = \left( 2,0,0\right)$, $\xi_{1}=\left(3,\frac{1}{2},+\frac{1}{2}\right)$ and $\xi_{2}=\left(4,0,0\right)$ are the set of quantum numbers $\left(N,S,S_{z}\right)$ labeling the eigenstates in $N$ particle sector; $S$ and $S_{z}$ being the total spin and its $z$-component, respectively.
\medskip
\label{tab:ksfe} \hfill \break}  
\end{table}
\begin{table}[!htb]
\resizebox{\linewidth}{!}{
\begin{tabular}{llllll}
\hline \hline
$\eta^{\prime}$ & Number on electron & Number on electron & Number on electron & Ground state & Ground state $\left|\lambda_{\xi_{2},0}^{\eta^{\prime}}\right\rangle$ of \\
     & on ${\bf H}_{eod}$ &  on ${\bf H}_{da}$ & on ${\bf H}_{1-site}$ & energy of ${\bf H}_{eod}$ & non-int ${\bf H}_{zbw}$ \\
\hline \hline
1 & 0 & 2 & 2 & - & $\left|\lambda_{\xi_{2},0}^{1}\right\rangle=\alpha^{\dag}_{a\uparrow}\alpha^{\dag}_{a\downarrow}d^{\dag}_{a\uparrow}d^{\dag}_{a\downarrow}\left|0\right\rangle$\\ 
2 & 1 & 2 & 1 & $\left|\lambda_{\xi_{0},0}^{eod}\right\rangle$ & $\left|\lambda_{\xi_{2},0}^{2}\right\rangle=\frac{1}{\sqrt{2}}d^{\dag}_{a\uparrow}d^{\dag}_{a\downarrow}\left(\alpha^{\dag}_{a\uparrow}\left|\lambda^{eod}_{1,\frac{1}{2},-\frac{1}{2},0}\right\rangle-\alpha^{\dag}_{a\downarrow}\left|\lambda^{eod}_{\xi_{0},0}\right\rangle\right)$\\ 
3 & 1 & 1 & 2 & $\left|\lambda_{\xi_{0},0}^{eod}\right\rangle$ & $\left|\lambda_{\xi_{2},0}^{3}\right\rangle=\frac{1}{\sqrt{2}}\alpha^{\dag}_{a\uparrow}\alpha^{\dag}_{a\downarrow}\left(d^{\dag}_{a\uparrow}\left|\lambda^{eod}_{1,\frac{1}{2},-\frac{1}{2},0}\right\rangle-d^{\dag}_{a\downarrow}\left|\lambda^{eod}_{\xi_{0},0}\right\rangle\right)$\\ 
4 & 2 & 1 & 1 & $\left|\lambda_{\xi,0}^{eod}\right\rangle$ &  $\left|\lambda_{\xi_{2},0}^{4}\right\rangle=\frac{1}{\sqrt{2}}\left(d^{\dag}_{a\uparrow}\alpha^{\dag}_{a\downarrow}+\alpha^{\dag}_{a\uparrow}d^{\dag}_{a\downarrow}\right)\left|\lambda_{\xi,0}^{eod}\right\rangle$ \\
5 & 2 & 2 & 0 & $\left|\lambda_{\xi,0}^{eod}\right\rangle$ & $\left|\lambda_{\xi_{2},0}^{5}\right\rangle=d^{\dag}_{a\uparrow}d^{\dag}_{a\downarrow}\left|\lambda_{\xi,0}^{eod}\right\rangle$\\ 
6 & 2 & 0 & 2 & $\left|\lambda_{\xi,0}^{eod}\right\rangle$ & $\left|\lambda_{\xi_{2},0}^{6}\right\rangle=\alpha^{\dag}_{a\uparrow}\alpha^{\dag}_{a\downarrow}\left|\lambda_{\xi,0}^{eod}\right\rangle$\\ 
7 & 3 & 1 & 0 & $\left|\lambda_{\xi_{1},0}^{eod}\right\rangle$ & $\left|\lambda_{\xi_{2},0}^{7}\right\rangle=\frac{1}{\sqrt{2}}\left(d^{\dag}_{a\uparrow}\left|\lambda^{eod}_{3,\frac{1}{2},-\frac{1}{2},0}\right\rangle-d^{\dag}_{a\downarrow}\left|\lambda^{eod}_{\xi_{1},0}\right\rangle\right)$ \\ 
8 & 3 & 0 & 1 & $\left|\lambda_{\xi_{1},0}^{eod}\right\rangle$ & $\left|\lambda_{\xi_{2},0}^{8}\right\rangle=\frac{1}{\sqrt{2}}\left(\alpha^{\dag}_{a\uparrow}\left|\lambda^{eod}_{3,\frac{1}{2},-\frac{1}{2},0}\right\rangle-\alpha^{\dag}_{a\downarrow}\left|\lambda^{eod}_{\xi_{1},0}\right\rangle\right)$\\ 
9 & 4 & 0 & 0 & $\left|\lambda^{eod}_{\xi_{2},0}\right\rangle$ & $\left|\lambda_{\xi_{2},0}^{9}\right\rangle=\alpha^{\dag}_{s\uparrow}\alpha^{\dag}_{s\downarrow}d^{\dag}_{s\uparrow}d^{\dag}_{s\downarrow}\left|0\right\rangle$\\ 
\hline \hline
 \end{tabular}}
\caption{ Possible ground states in the non-interacting case. Where $\lambda_{\xi_{0},0}^{eod}=\frac{1}{2}\left( \varepsilon+\varepsilon_{F}+t-R_{0}^{\prime}\right)$, $\lambda_{\xi,0}^{eod}=\varepsilon+\varepsilon_{F}+t-R_{0}^{\prime}$, $\lambda_{\xi_{1},0}^{eod}=\frac{1}{2}\left[3\left(\varepsilon+\varepsilon_{F}+t\right)-R^{\prime}_{0} \right]$, $\lambda_{\xi_{2},0}^{eod}=2\left(\varepsilon+\varepsilon_{F}+t\right)$ and $R_{0}^{\prime}=\sqrt{\left(\varepsilon-\varepsilon_{F}+t\right)^{2}+16V^{2}}$. The eigenstates $\left|\lambda_{\xi_{2},0}^{\eta^{\prime}}\right\rangle$ can be obtained easily with the help of eigenstates of ${\bf H}_{eod}$ corresponding to the eigenvalues $\lambda_{\xi_{0},0}^{eod}$, $\lambda_{\xi,0}^{eod}$, $\lambda_{\xi_{1},0}^{eod}$ and $\lambda_{\xi_{2},0}^{eod}$. The symbols $\xi_{0}= \left(1,\frac{1}{2},+\frac{1}{2}\right)$, $\xi = \left( 2,0,0\right)$, $\xi_{1}=\left(3,\frac{1}{2},+\frac{1}{2}\right)$ and $\xi_{2}=\left(4,0,0\right)$ are the set of quantum numbers $\left(N,S,S_{z}\right)$ labeling the eigenstates in $N$ particle sector; $S$ and $S_{z}$ being the total spin and its $z$-component, respectively. 
\hfill \break}
\label{tab:gsnonint4e}  
\end{table}
\\
\indent
In the non-interacting case ({\it i.e.} $U=0$ and $g=0$) the possible ground states are listed in the Table \ref{tab:gsnonint4e}. The explicit expression for them can be easily found, for the case when Fermi levels in the leads are set at $\varepsilon_{F}=0$ and the dot levels below it $\varepsilon < \varepsilon_{F}$, the ground state in this case corresponds to the eigenvalue $\lambda_{\xi_{2},0}^{5}$ is given by
\begin{small}
\begin{eqnarray}
\left|\lambda_{\xi_{2},0}^{5}\right\rangle &=& C^{1}_{0}\left|D\right\rangle+C^{2}_{0}\alpha^{\dag}_{s\uparrow}\alpha^{\dag}_{s\downarrow}\left[\frac{1}{2}\left(\left|i\right\rangle+\left|\bar{i}\right\rangle \right)- \frac{1}{\sqrt{2}}\left|\Xi\right\rangle\right]
\nonumber \\
&+&
C^{3}_{0}\left[
\frac{1}{\sqrt{2}}\left(c^{\dag}_{1\uparrow}\alpha^{\dag}_{s\downarrow}+\alpha^{\dag}_{s\uparrow}c^{\dag}_{1\downarrow}\right)\left|\bar{i}\right\rangle+
\frac{1}{\sqrt{2}}\left(c^{\dag}_{2\uparrow}\alpha^{\dag}_{s\downarrow}
+\alpha^{\dag}_{s\uparrow}c^{\dag}_{2\downarrow}\right)\left|i\right\rangle
\right]
\label{nonintgs}
\end{eqnarray}
\end{small}
where $C_{0}^{1}=\pm\frac{1}{2R_{0}^{\prime}}\left(\varepsilon-\varepsilon_{F}+t-R_{0}^{\prime} \right)$, $C_{0}^{2}=\mp\frac{1}{2R_{0}^{\prime}}\left(\varepsilon-\varepsilon_{F}+t+R_{0}^{\prime} \right)$ and $C_{0}^{3}=\pm 2\sqrt{2}\frac{V}{R_{0}^{\prime}}$. 
The spin-spin correlation between the dots in this ground state becomes 
\begin{eqnarray}
\left\langle \lambda_{\xi_{2},0}^{5}\right| {\bf S_{1}\cdot S_{2}}\left|\lambda_{\xi_{2},0}^{5}\right\rangle = -\frac{3}{8}\left|C^{2}_{0}\right|^{2}
\label{nonintspincorr}
\end{eqnarray}
\indent
If the spins ${\bf S_{1}}$ and ${\bf S_{2}}$ associated with the dots are considered as the spins decoupled from the leads, the two dots can form a singlet ($\tilde{S}=0$) or a triplet ($\tilde{S}=1$) and the spin-spin correlation 
${\bf S_{1}.S_{2}}=\frac{1}{2}\left(\tilde{S}^{2}-S_{1}^{2}-S_{2}^{2} \right)$ is given by
\begin{eqnarray}
\bf \left\langle S_{1}.S_{2}\right\rangle=\left\lbrace
\begin{array}{lr}
-\frac{3}{4} & \,\, \tilde{S}=0\\
+\frac{1}{4} & \,\, \tilde{S}=1 .
\end{array}
\right.
\label{twospin}
\end{eqnarray}
Thus as the limiting case spin-spin correlation between the dots for our four-site half-filled case are bounded as $-\frac{3}{4} \le \bf \left\langle S_{1}.S_{2}\right\rangle  \le +\frac{1}{4}$.
\section{Numerical Results}
\label{numbers}
The numerical calculations are done using analytical expressions for the eigenstates in Table {\ref{tab:gsfe}} and spin-spin correlation in Table {\ref{tab:ksfe}} for the half-filled case. The hybridization of the dots with the leads $V$ is usually kept as weak as possible so that the number of confined electrons are prevented from strong fluctuations \cite{cronenwett}. We fixed the Fermi energy of the leads at $\varepsilon_{F}=0\ V$ and dot energies at $\varepsilon=-5V$ i.e. below the Fermi level so as to further prevent the fluctuations in the confined electron number, taking the hybridization $V$ as the smallest parameter, the unit of energy. 
In Fig. \ref{corrg0} we plot spin-spin correlation between the dots $ \bf {\left\langle S_{1}\cdot S_{2} \right\rangle}$ as a function of interdot tunneling matrix-element $t$ and ondot Coulomb interaction $U$ at fixed value of the interdot Coulomb interaction $g=0\ V$. The spin-spin correlation between the dots can be classified into two regions identified as having ferromagnetic $\left( \bf {\left\langle S_{1}\cdot S_{2} \right\rangle}>0\right)$ and antiferromagnetic $\left( \bf {\left\langle S_{1}\cdot S_{2} \right\rangle}<0\right)$ correlations. 
It is observed that the ferromagnetic correlation between the dots takes place for $U\gg\left|\varepsilon\right| \ge t$. The ferromagnetic correlation attains its maximum value ${\bf \left\langle S_{1}\cdot S_{2}\right\rangle}\approx\frac{1}{4}$ for small values of interdot tunneling matrix-element $t\sim V$.
Different type of spin-spin correlation between the dots in the half-filled case, listed in Table \ref{tab:ksfe}, can be understood with the help of corresponding many-body ground state in the total spin $S=0$ subspace, listed in Table \ref{tab:gsfe}. Ferromagnetic correlation between the dots takes place when each dot has an average occupancy of one-electron with parallel spins. 
 The other two-electrons are present on the leads with their spins anti-parallel to the dot spins so as to give total spin $S=0$. Such a configuration for the state is favored when ondot Coulomb interaction is large $U\gg\left|\varepsilon\right|> t$ so as to avoid double occupancy on the dots. In the ferromagnetic region shown in Fig. \ref{corrg0}, the average occupancy of the two dots in the corresponding region is nearly one $\left\langle n_{i}\right\rangle\approx1$, as can be seen from Fig. \ref{avgg0} for dot-1 (since the two dots are identical, we give occupation number of dot-1 only). 
Figure \ref{gsg0} shows values of integer $\eta$ corresponding to one of the possible ground states listed in Table \ref{tab:gsfe}, in $U$-$t$ parameter space. For the ferromagnetic correlation, the ground state of the system is found to correspond to $\eta=3$ and
the sign of spin-spin correlation $\left\langle{\bf S_{1}\cdot S_{2}}\right\rangle$ in this state is positive, as seen in Table \ref{tab:ksfe}.
\begin{figure}[!htb]
\centering
\subfigure[${\bf \left\langle S_{1}\cdot S_{2}\right\rangle}$]
{\includegraphics[width=0.30\linewidth, height=9em]{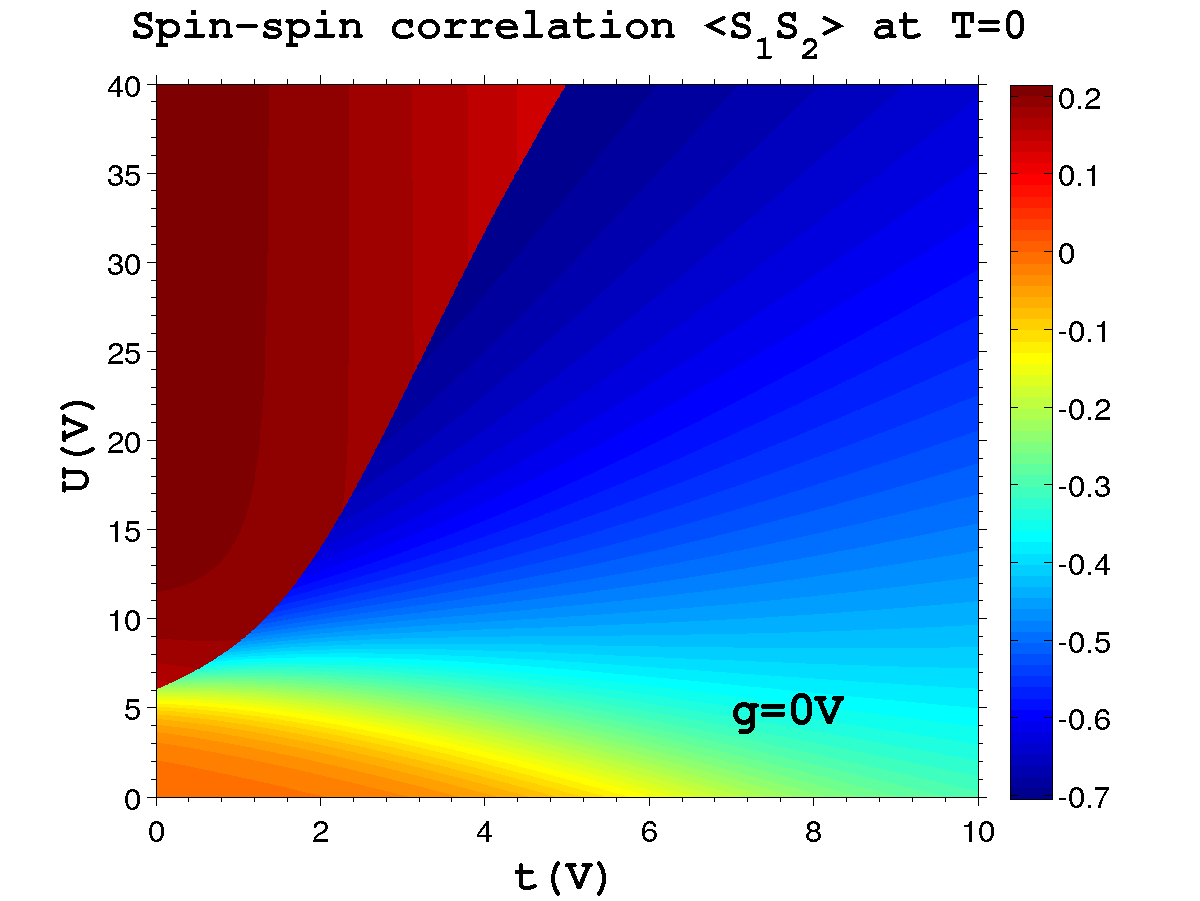}\label{corrg0}}
{\hglue 4mm}
\subfigure[Avg. occupation $\left\langle n_{1}\right\rangle$]
{\includegraphics[width=0.30\linewidth, height=9em]{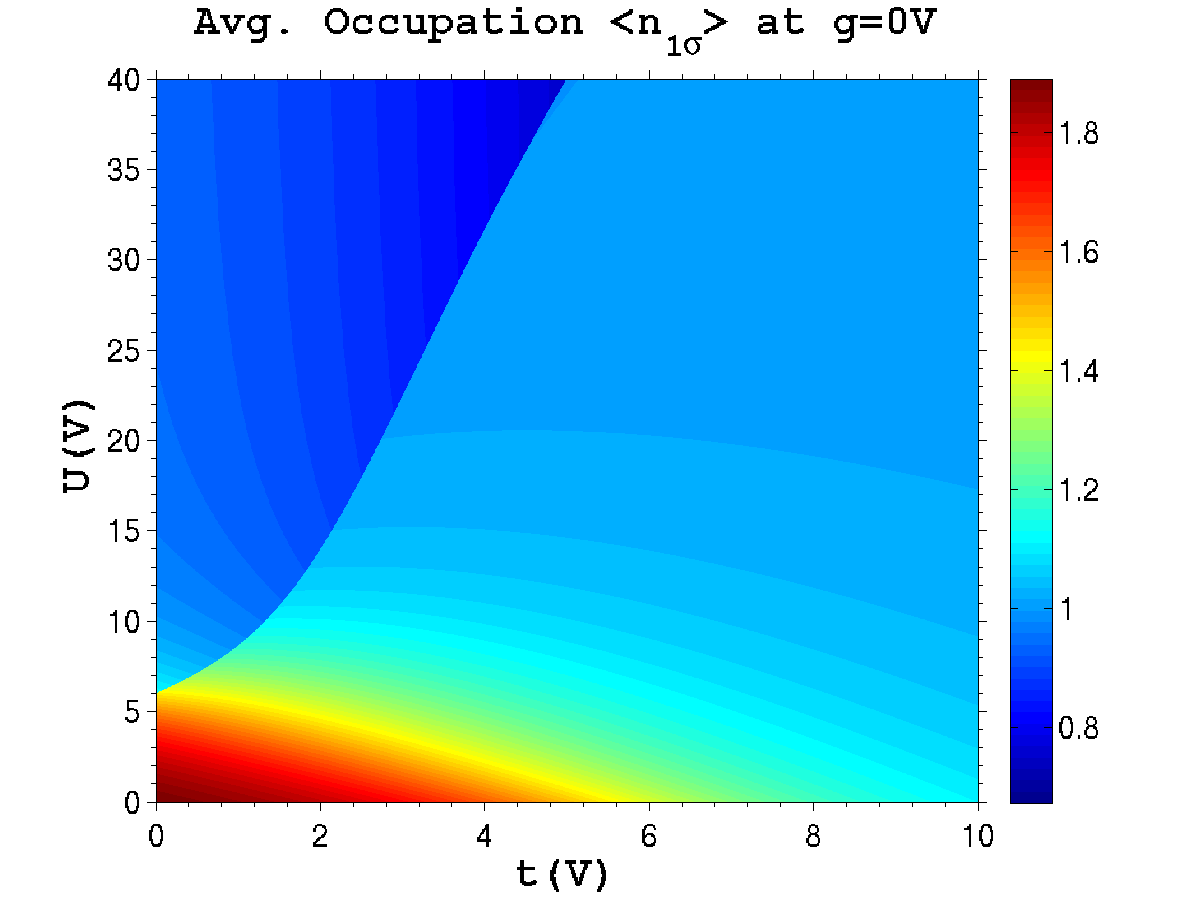}\label{avgg0}}
{\hglue 4mm}
\subfigure[Ground state $\left|\lambda^{4,\eta}_{\xi_{2}}\right\rangle$]
{\includegraphics[width=0.30\linewidth, height=9em]{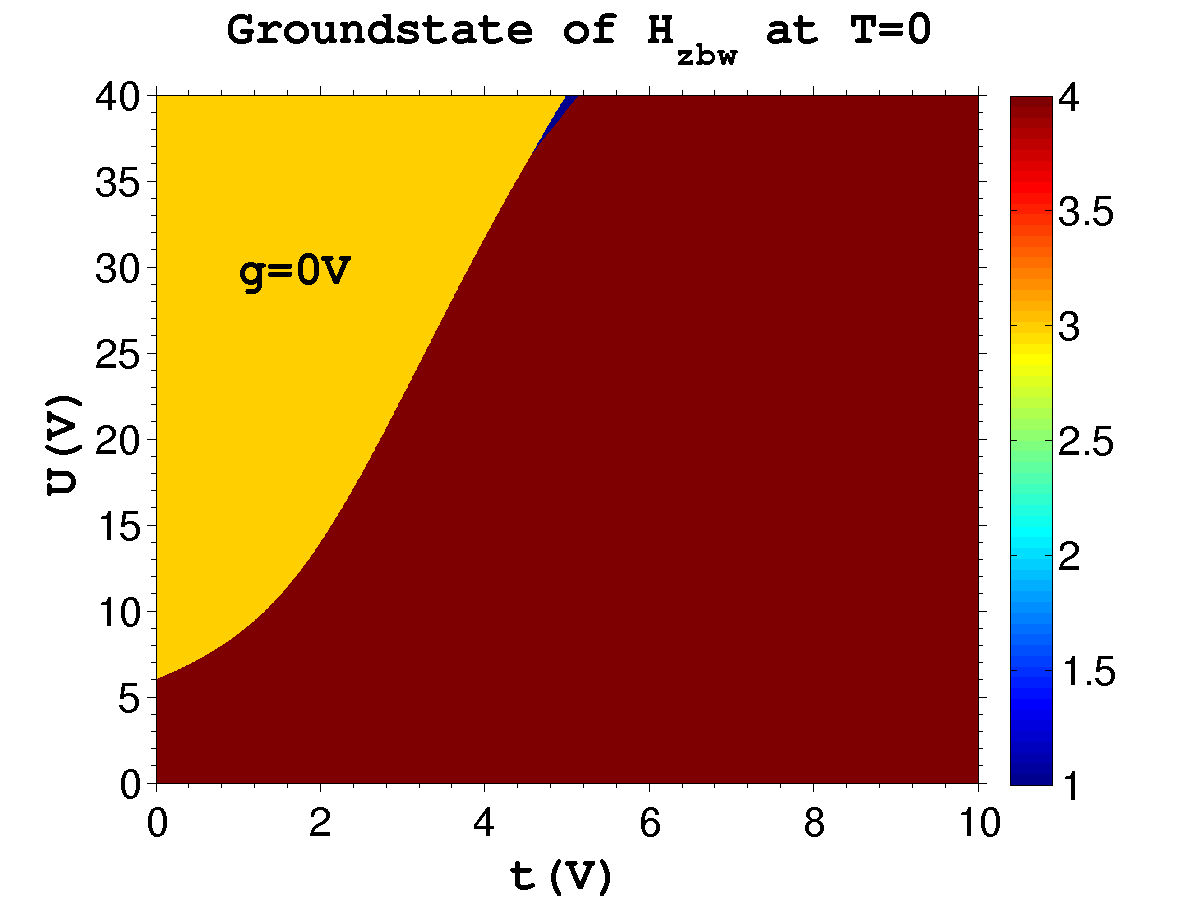}\label{gsg0}}
\caption{ For the half-filled case with the Fermi energy fixed at $\varepsilon_{F}=0$, the dot energies at $\varepsilon=-5V$ and the interdot Coulomb interaction at $g=0V$; in the unit of the hybridization parameter $V$. In (a) We show the spin-spin correlation ${\bf \left\langle S_{1}\cdot S_{2}\right\rangle}$ between the dots for the half-filled case as a function of interdot tunneling matrix-element $t$ and ondot Coulomb interaction $U$. The possible values lie between $-\frac{3}{4}$ to $+\frac{1}{4}$. Positive values of ${\bf \left\langle S_{1}\cdot S_{2}\right\rangle}>0$ signifies the ferromagnetic correlation whereas the negative values ${\bf \left\langle S_{1}\cdot S_{2}\right\rangle}<0$ the antiferromagnetic correlation between the dots. (b) We show the average occupation of the dot-1 $\left\langle n_{1}\right\rangle$ as a function of interdot tunneling matrix-element $t$ and ondot Coulomb interaction $U$, can have values between $0$ to $2$. (c) We show the ground state of the half-filled system as a function of interdot tunneling matrix-element $t$ and ondot Coulomb interaction $U$. The integer value of $\eta$ identifies one of the possible ground states listed in Table \ref{tab:ksfe}. In the  figure above, yellow corresponds to $\eta =3$ and brown to $\eta =4$. 
\label{fig:tUg0}\hfill \break}
\end{figure}
The ferromagnetic correlation can have a maximum value of $\frac{1}{4}$ weighted by the coefficient $C^{5,5}_{\xi_{1}}$ in eq. (\ref{gseta3}) determined by the system parameters. The coefficient $C^{5,5}_{\xi_{1}}$ in the ground state in eq. (\ref{gs3}) is the probability amplitude of  the state  $\frac{1}{\sqrt{6}}\left[
\left(\alpha^{\dag}_{s\uparrow}\alpha^{\dag}_{a\downarrow}+\alpha^{\dag}_{s\downarrow}\alpha^{\dag}_{a\uparrow}\right) \left|\Theta \right\rangle+\sqrt{2}\alpha^{\dag}_{a\uparrow}\alpha^{\dag}_{s\uparrow}\left|\bar{\sigma}\right\rangle+\sqrt{2}\alpha^{\dag}_{a\downarrow}\alpha^{\dag}_{s\downarrow}\left|\sigma\right\rangle 
\right]$, which has spins on the dots $\left|\Theta \right\rangle$, $\left|\bar{\sigma}\right\rangle$ and $\left|\sigma\right\rangle$ coupled to form a triplet.
\\
 In $U$-$t$ parameter space as can be seen from Fig. \ref{corrg0}, most of the region corresponds to the antiferromagnetic correlation between the dots $(\left<{\bf S_{1}\cdot S_{2}}\right> <0)$. The average occupancies $\left< n_{i}\right>$ of the dots varies between 1 and 2 as observed in Fig. \ref{avgg0}. For large values of the ondot Coulomb interaction $U>>|\varepsilon|$ the dots are singly occupied $\left< n_{i}\right>\sim 1$ and for small values $U\le |\varepsilon|$ the dots can have double occupancies $\left< n_{i}\right>\sim 2$. The ground state corresponds to $\eta=4$ in Table \ref{tab:gsfe} as evident from Fig. \ref{gsg0}. From its explicit expression given in the eq. (\ref{gs4}) it can be seen that the contribution to spin-spin correlation $\left< {\bf S_{1}\cdot S_{2}}\right>$ comes only from first term corresponding to the the coefficient $C_{\xi_{2}^{1,1}}$ as given in the Table {\ref{tab:ksfe}}. The basis state  $\alpha^{\dag}_{s\uparrow}\alpha^{\dag}_{s\downarrow}\left|\Xi\right\rangle $ with probability amplitude $C_{\xi_{2}^{1,1}}$ shows that the electrons on the dots form a singlet $\left|\Xi \right\rangle$. The spin-spin correlation can have its maximum value $\left<{\bf S_{1}\cdot S_{2}}\right>\approx -\frac{3}{4}$ in $U-t$ parameter space for $U\gg \left| \varepsilon \right|$ and $t\gg V$.
 \\
\indent
{\bf Interdot tunneling matrix-element $t\approx 0$:} The two dots in parallel geometry are correlated, directly through the tunneling matrix-element $t$ and indirectly via leads through the hybridization parameter $V$. Due to this fact the model exhibits correlation between the dots even for vanishing interdot tunneling matrix-element $t\approx 0$, as can be seen from Fig. {\ref{corrg0}}. In this case the ondot Coulomb interaction $U$ plays a key role in controlling the occupancies on the dots resulting in ferromagnetic or antiferromagnetic correlation between the dots. For $U\leq \left|\varepsilon\right|$, the dots can possibly be doubly occupied as the average occupation number on each dot takes values $1\le\left\langle n_{i}\right\rangle\le 2$, as can be seen from Fig. \ref{avgg0}. The corresponding correlation between the dots is antiferromagnetic as can be seen from Fig. {\ref{corrg0}}. This can be understood by  considering a perturbation scheme for a three  particle state \cite{rallub}. If a three particle state contains two-electrons, one on each dot with anti-parallel spins and the third electron on leads; this enables one of the dot electrons to transfer to the leads and then to the other dot (indirect exchange) through the hybridization parameter $V$. From Fig. {\ref{gsg0}}, we find that the ground state corresponds to the state $\eta=4$ in Table {\ref{tab:gsfe}} with the corresponding sign of ${\bf \left\langle S_{1}\cdot S_{2}\right\rangle}$ given in Table {\ref{tab:ksfe}} as negative, signifying that the correlation is antiferromagnetic. \\
\indent
As the ondot Coulomb interaction becomes large $U\gg \left|\varepsilon\right|$, the dots exhibits ferromagnetic correlation between them. This can again be understood through perturbations considering a three particle state. If a three particle state contains two-electrons, one on each dot (double occupancy avoided due to large $U$) and the third electron on the leads. In order to lower the ground state energy, the spins of the electrons on the dots must be aligned parallel (for ferromagnetic correlation to occur) and aligned antiparallel with respect to the lead electrons. The fourth electron is aligned appropriately so that the total spin of the four-electron system is zero, $S=0$; such a configuration is clearly seen in the state $\left|2\right\rangle_{3e}$ obtained in eq. (\ref{gseta3}).
\\
\indent
{\bf Non-interacting case:} In the absence of ondot and interdot interactions {\it i.e.} $U=0$ and $g=0$, the spin-spin correlation between the dots disappears with ${\bf \left\langle S_{1}\cdot S_{2}\right\rangle}\approx 0$ for small values of interdot tunneling matrix-element $t\sim V$ as seen in Fig. {\ref{coulombint}. However, for large values of the interdot tunneling matrix-element $t\gg V$, the dots exhibit antiferromagnetic correlation between them as seen in Fig. \ref{coulombint}. This behavior of antiferromagnetic correlation is similar for $U<\left|\varepsilon\right|$, as shown for three values of ondot Coulomb interaction $U=0.5V,1V,2V$ in Fig. \ref{coulombint}. In this situation, two electrons with opposite spins can reside on a dot (Pauli exclusion principle) and the interdot tunneling matrix-element $t$ may cause one of the electrons to transfer to the other dot; occupied by an electron with opposite spin. With increasing interdot tunneling matrix-element $t$, the spin-spin correlation between the dots attains a maximum value of ${\bf \left\langle S_{1}\cdot S_{2}\right\rangle}= -\frac{3}{8}$, as can be readily verified by explicit expression for  ${\bf \left\langle S_{1}\cdot S_{2}\right\rangle}$ for the half-filled case obtained from eq. (\ref{nonintspincorr}). This can be clearly observed from Fig. \ref{coulombint} or from Fig. {\ref{corrg0}}.
\begin{figure}[!htb]
\centering
\subfigure[At different $U$]
{\includegraphics[width=0.45\linewidth]{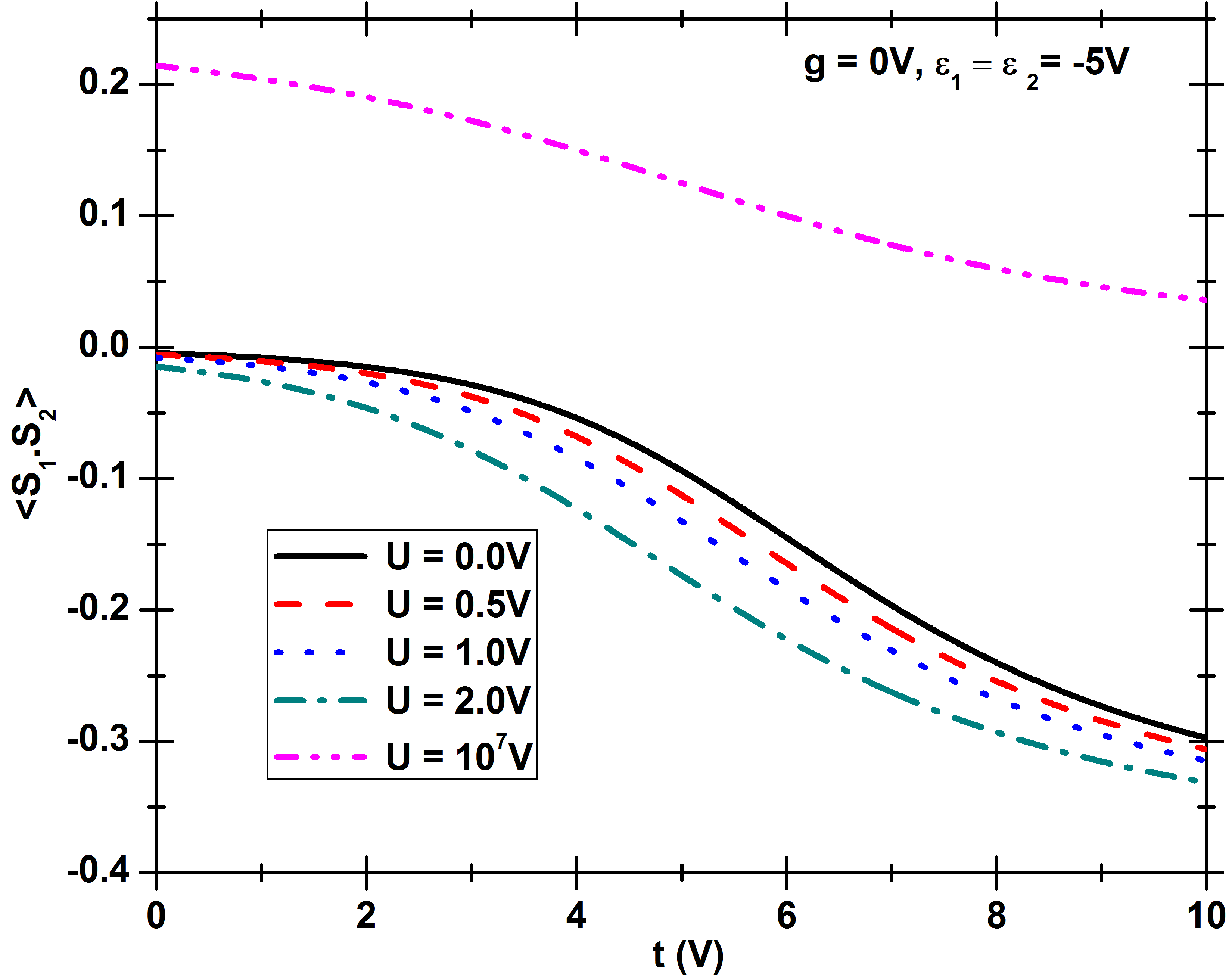}\label{coulombint}}
{\hglue 4mm}
\subfigure[At different $t$]
{\includegraphics[width=0.45\linewidth]{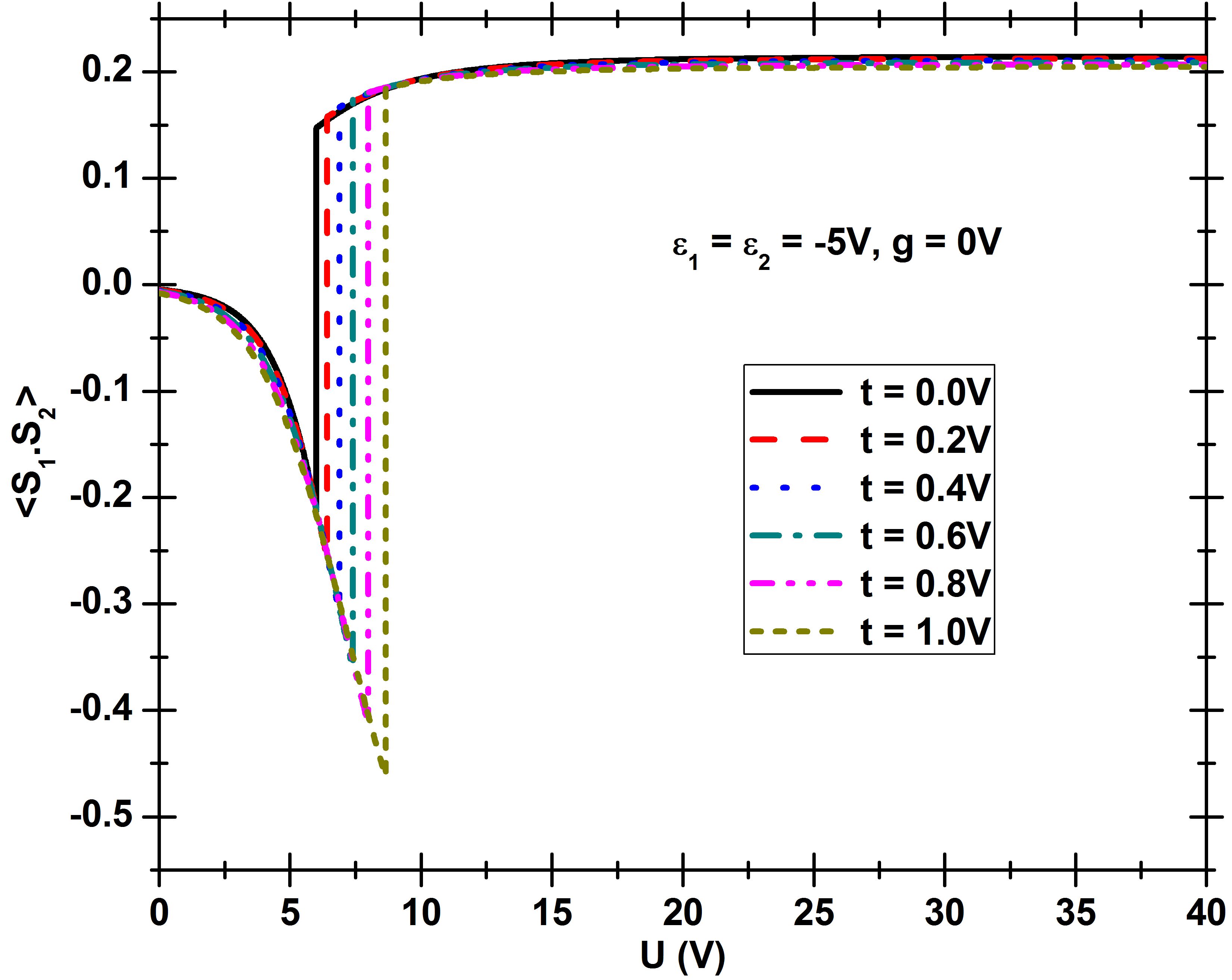}\label{difft}}
\caption{ Zero temperature spin-spin correlation ${\bf <S_{1}\cdot S_{2}>}$ between the dots in (a) as a function of the interdot tunneling matrix-element $t$ for Non-Interacting case $U<\left|\varepsilon\right|$ and infinite $U\rightarrow \infty$ case. The plots correspond to five different values of ondot Coulomb interaction: $U=0.0-$solid line; $U=0.5V-$dashed line; $U=1V-$dotted line; $U=2V-$dash-dotted line; $U=10^{7}V-$dash double dotted line. (b) as a function of the ondot Coulomb interaction $U$. The plots correspond to six different values of interdot tunneling matrix-element $t$: $t=0-$Solid line; $t=0.2V-$dashed line; $t=0.4V-$dotted line; $t=0.6V-$dash-dotted line; $t=0.8V-$dash double dotted line and $t=1V-$short-dashed line.\hfill \break
\label{fig:nonintUinf} \hfill \break}
\end{figure} 
In Fig. {\ref{coulombint}}, we also have plotted the spin-spin correlation ${\bf \left\langle S_{1}\cdot S_{2}\right\rangle}$ for very large value of ondot coulomb interaction $U=10^{7}\sim \infty$. As now the dots can only be singly occupied, it exhibit ferromagnetic correlation for any value of interdot tunneling matrix element $t$. The maximum value is found to be ${\bf \left\langle S_{1}\cdot S_{2}\right\rangle}\approx \frac{1}{4}$ cand can be verified through analytical value calculated in eq. ({\ref{corrUinf}}) in $U\rightarrow\infty$ limit.
In Fig. {\ref{difft}}, we have plotted the spin-spin correlation ${\bf \left\langle S_{1}\cdot S_{2}\right\rangle}$ between the dots as a function of ondot Coulomb interaction $U$ for six different values of interdot tunneling matrix-element $t=0.0,0.2V,0.4V,0.6V,0.8V,1.0V$ in the absence of interdot interaction $g=0$. It is observed that the interdot tunneling matrix-element $t$ and the ondot Coulomb interaction $U$ has a critical dependency {\it i.e.} for a given value of $t$ there is a critical value of $U$ leading to the transition from antiferromagnetic correlation to ferromagnetic correlation.
\begin{figure}[!htb]
\centering
\subfigure[${\bf \left\langle S_{1}\cdot S_{2}\right\rangle}$ at $g=5$]
{\includegraphics[width=0.30\linewidth, height=9em]{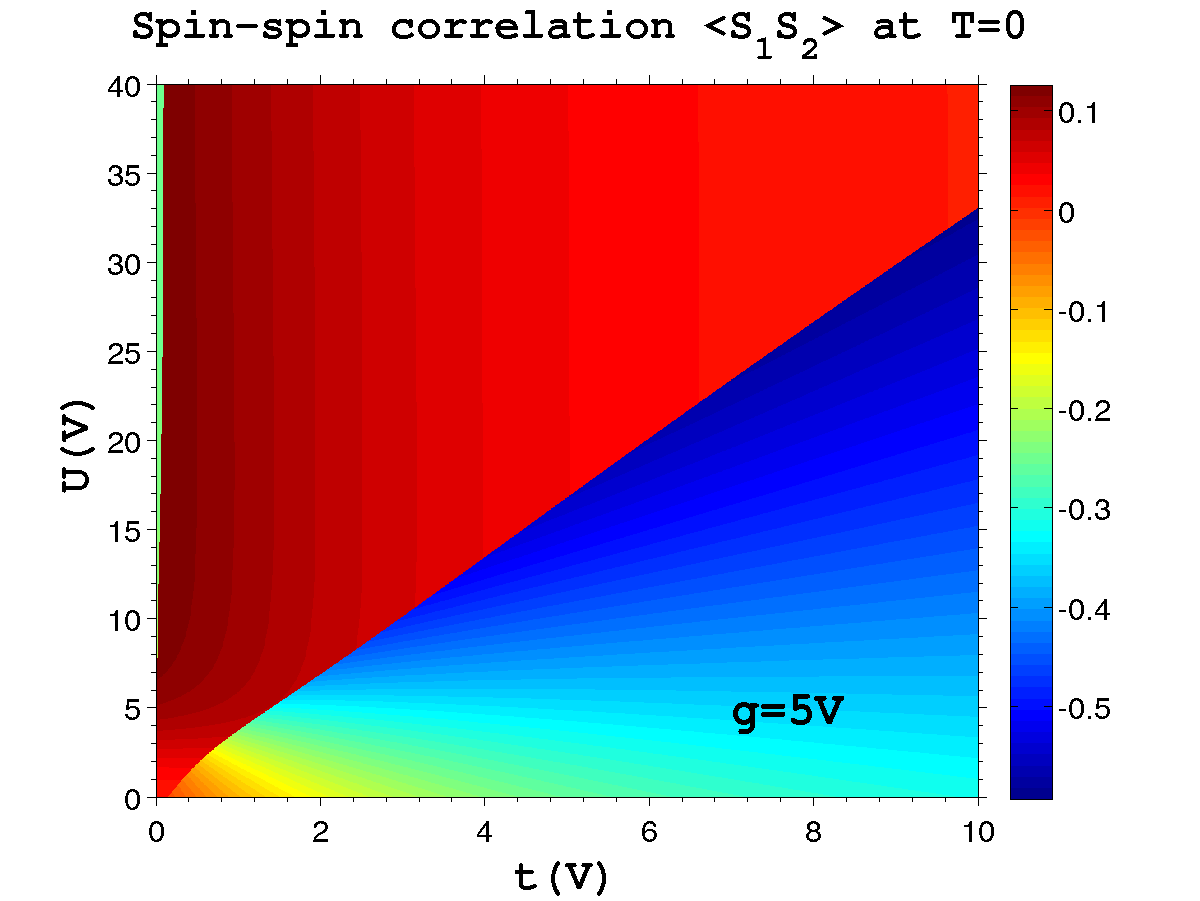}\label{corrg5}}
{\hglue 4mm}
\subfigure[Avg. occupation $\left\langle n_{1\sigma}\right\rangle$]
{\includegraphics[width=0.30\linewidth, height=9em]{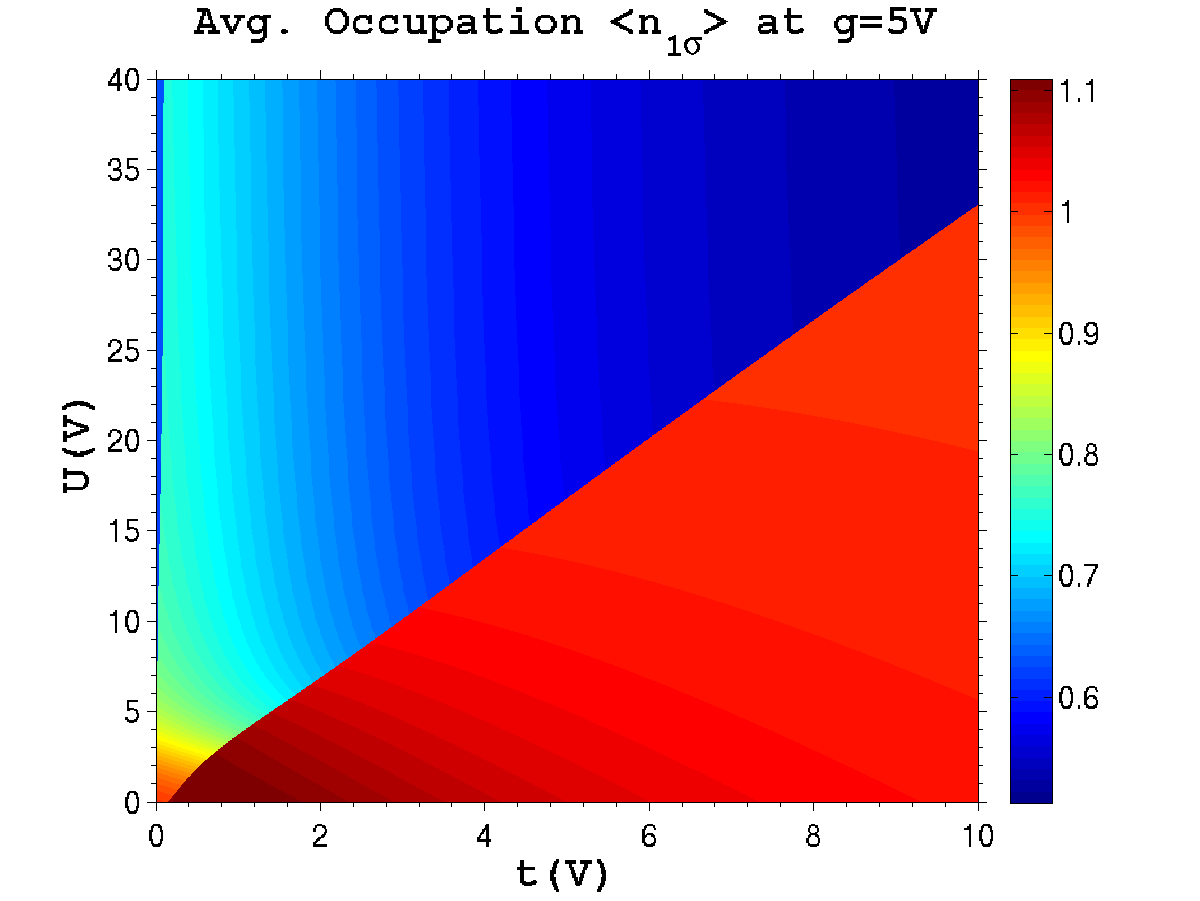}\label{avgg5}}
{\hglue 4mm}
\subfigure[Ground state $\left|\lambda^{4,\eta}_{\xi_{2}}\right\rangle$ ]
{\includegraphics[width=0.30\linewidth, height=9em]{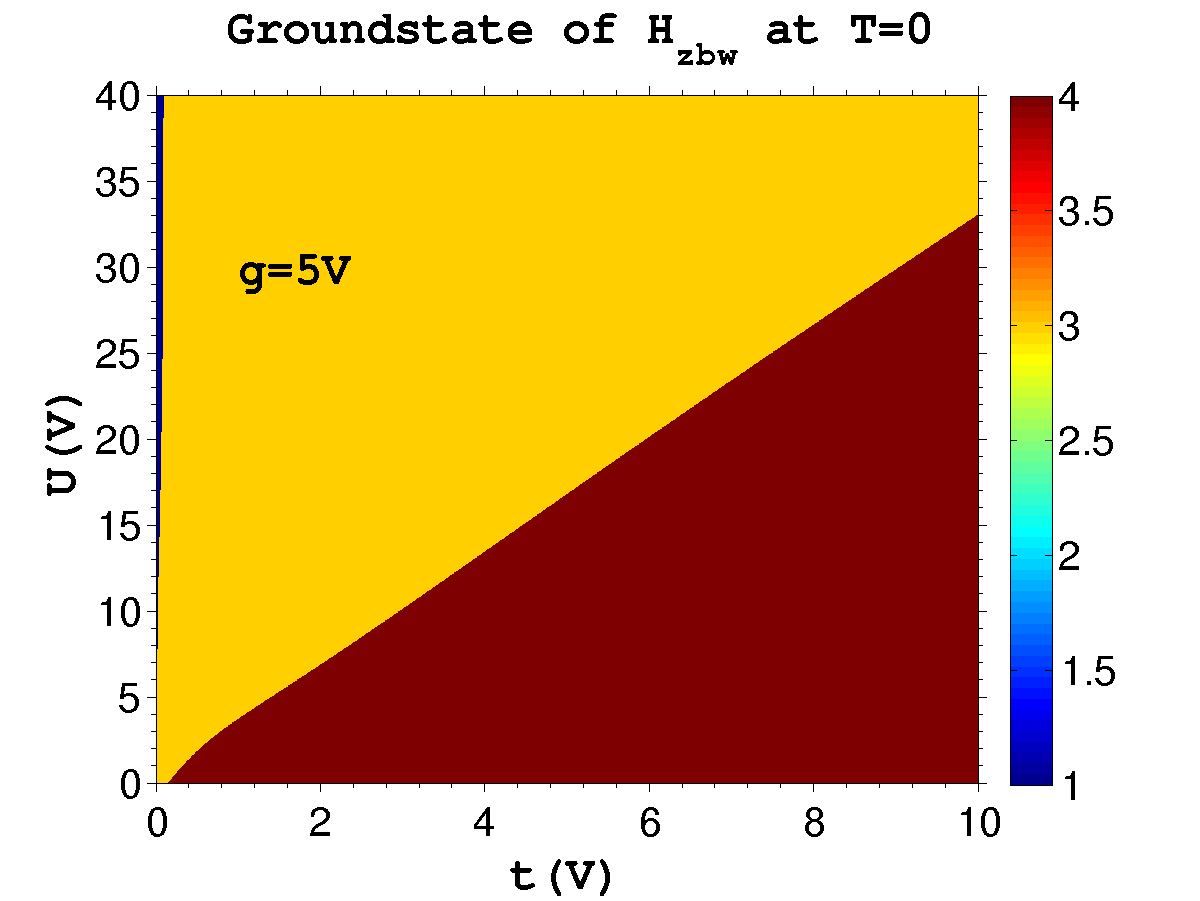}\label{gsg5}}
\caption{ For the half-filled case with the Fermi energy fixed at $\varepsilon_{F}=0$, the dot energies at $\varepsilon=-5V$ and the interdot Coulomb interaction at $g=5V$; in the unit of the hybridization parameter $V$. In (a) We show the spin-spin correlation ${\bf \left\langle S_{1}\cdot S_{2}\right\rangle}$ between the dots for the half-filled case as a function of interdot tunneling matrix-element $t$ and ondot Coulomb interaction $U$. The possible values lie between $-\frac{3}{4}$ to $+\frac{1}{4}$. Positive values of ${\bf \left\langle S_{1}\cdot S_{2}\right\rangle}>0$ signifies the ferromagnetic correlation whereas the negative values ${\bf \left\langle S_{1}\cdot S_{2}\right\rangle}<0$ the antiferromagnetic correlation between the dots. (b) We show the average occupation of the dot-1 $\left\langle n_{1}\right\rangle$ as a function of interdot tunneling matrix-element $t$ and ondot Coulomb interaction $U$, can have values between $0$ to $2$. (c) We show the ground state of the half-filled system as a function of interdot tunneling matrix-element $t$ and ondot Coulomb interaction $U$. The integer value of $\eta$ identifies one of the possible ground states listed in Table \ref{tab:ksfe}. In the figure above, blue corresponds to $\eta =1$, yellow to $\eta =3$ and brown to $\eta =4$.
\label{fig:tUg5}\hfill \break}
\end{figure}
\\
In Fig. {\ref{corrg5}}, we have plotted the spin-spin correlation ${\bf \left\langle S_{1}\cdot S_{2}\right\rangle}$ between the dots for the half-filled case as a function of interdot tunneling matrix-element $t$ and the ondot Coulomb interaction $U$ at a fixed value of interdot Coulomb interaction $g=5$. It is observed that the ferromagnetic correlation in the $U-t$ parameter space corresponding to ${\bf \left\langle S_{1}\cdot S_{2}\right\rangle}>0$ occupies large region of space as compared to the antiferromagnetic region ${\bf \left\langle S_{1}\cdot S_{2}\right\rangle}<0$. For small values of interdot tunneling matrix-element $t\lesssim V$, the dots exhibit antiferromagnetic correlation even for large values of ondot Coulomb interaction $U\gg\left|\varepsilon\right|$ unlike the $g=0$ case in Fig. {\ref{corrg0}} where the dots are correlated ferromagnetically. The interdot Coulomb interaction $g$ restricts the charge transfer between the dots due to the tunneling matrix-element $t$ and also renormalizes the ondot Coulomb interactions on the two dots. This brings into play the indirect exchange interaction between the dots via the leads through the hybridization parameter $V$. 
The ground state of the system in this situation corresponds to $\eta=1$ shown in Fig.  {\ref{gsg5}}. From the explicit expression given in Table-{\ref{tab:ksfe}} for the spin-spin correlation between the dots calculated using the ground state corresponding to $\eta=1$ in eq. (\ref{gs1}), it is seen that the antiferromagnetic correlation depends on the coefficient $C_{\xi}^{1,3}$. The coefficient $C_{\xi}^{1,3}$ is the probability amplitude of the state $\alpha^{\dag}_{a\uparrow}\alpha^{\dag}_{a\downarrow}\left|\Xi\right\rangle$ clearly showing that the electrons on the dots form a singlet $\left|\Xi\right\rangle$. From eq. (\ref{gseta1}) it is observed that the coefficient $C_{\xi}^{1,3}$ depends on the hybridization parameter $V$ allowing antiferromagnetic correlation to take place via leads. From Fig. {\ref{corrg5}} it is observed that the critical dependency of the interdot tunneling matrix-element $t$ on ondot Coulomb interaction $U$ causes alternate change of spin-spin correlation between the dots from antiferromagnetic to ferromagnatic then again to antiferromagnetic. Consequently, the ground state of the system changes from $\left|\lambda_{\xi_{2}}^{4,1}\right\rangle$ to $\left|\lambda_{\xi_{2}}^{4,3}\right\rangle$ and then to $\left|\lambda_{\xi_{2}}^{4,4}\right\rangle$ as shown in Fig {\ref{gsg5}}. The corresponding occupancies of the dots is nearly one $<n_{i}>\approx 1$ as shown for dot-1 in Fig. {\ref{avgg5}}. 
\begin{figure}[!htb]
\centering
\subfigure[${\bf \left\langle S_{1}\cdot S_{2}\right\rangle}$ at $g=10$]
{\includegraphics[width=0.30\linewidth, height=9em]{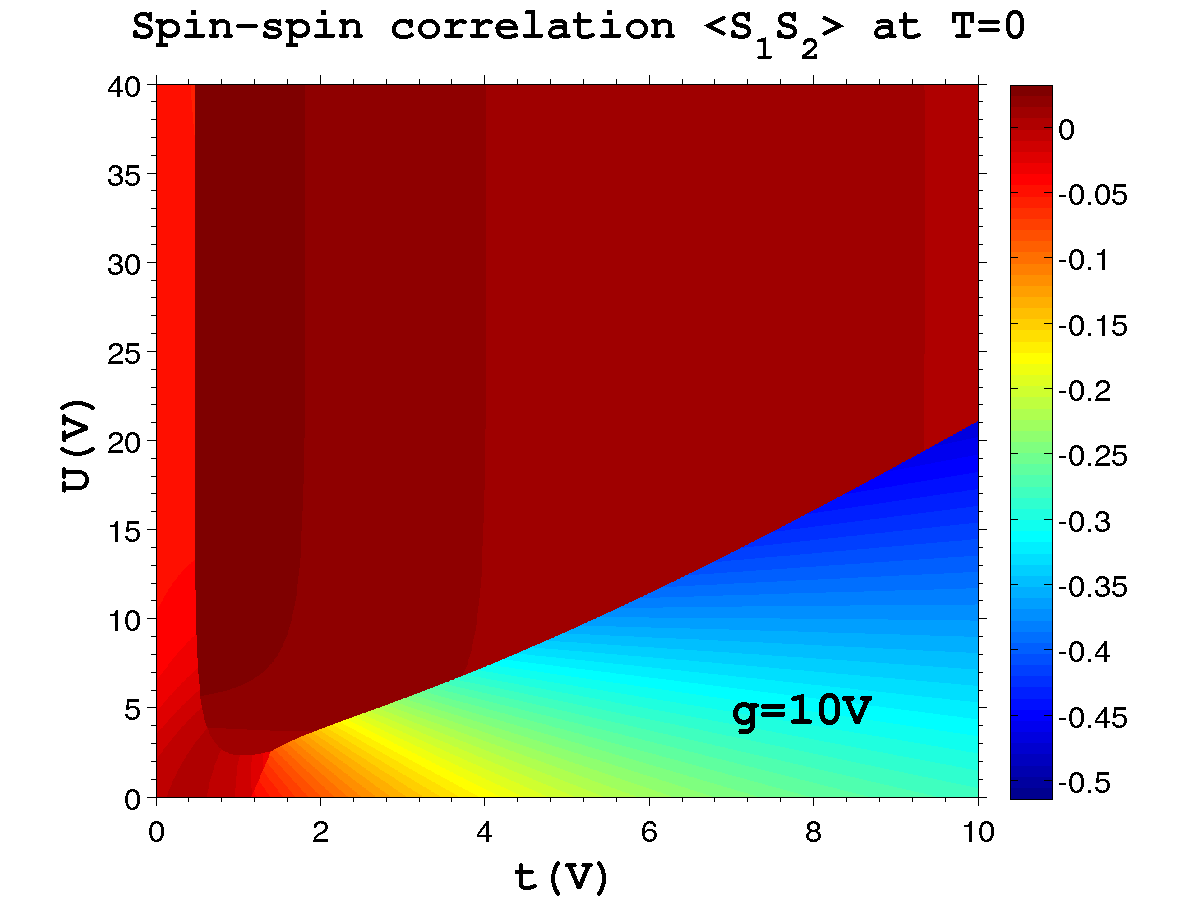}\label{corrg10}}
{\hglue 4mm}
\subfigure[Avg. occupation $\left\langle n_{1\sigma}\right\rangle$]
{\includegraphics[width=0.30\linewidth, height=9em]{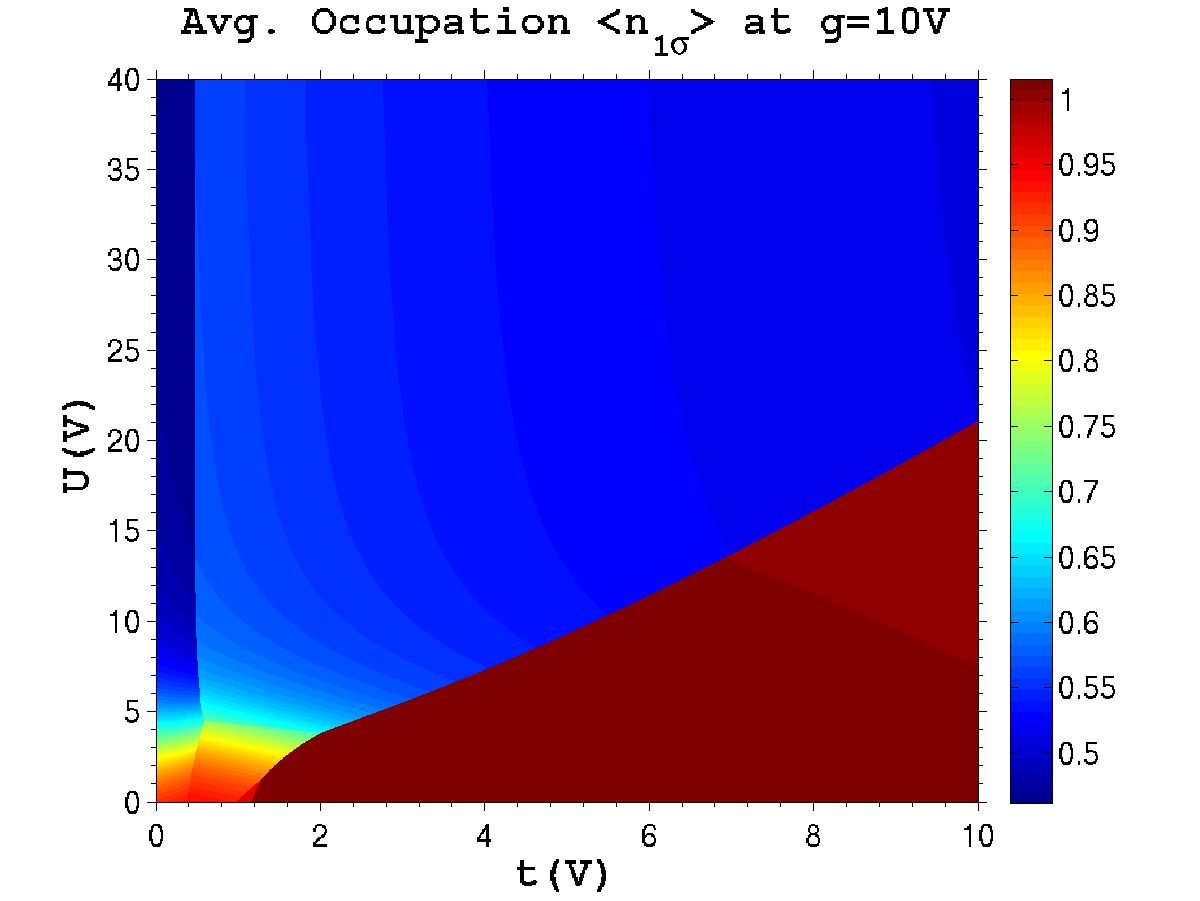}\label{avgg10}}
{\hglue 4mm}
\subfigure[Ground state $\left|\lambda^{4,\eta}_{\xi_{2}}\right\rangle$ ]
{\includegraphics[width=0.30\linewidth, height=9em]{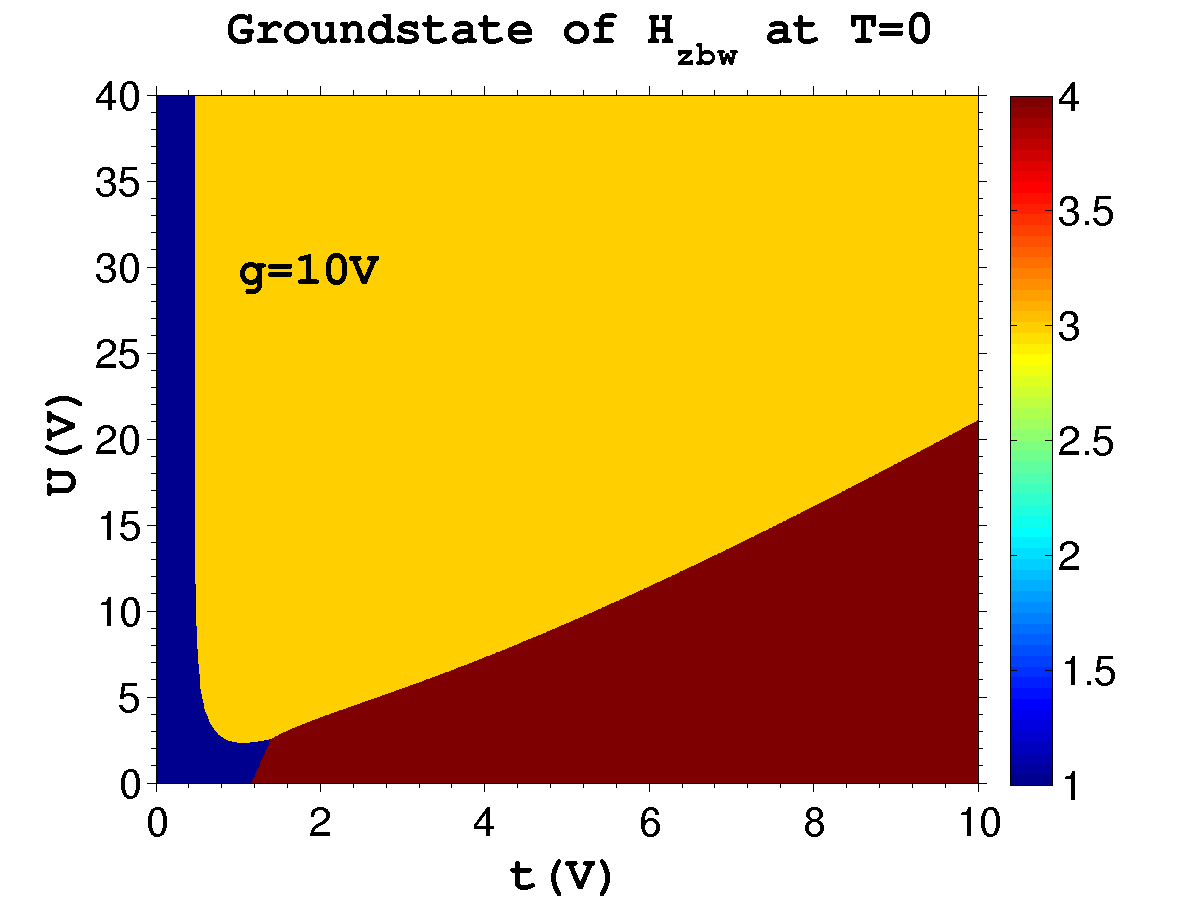}\label{gsg10}}
\caption{For the half-filled case with the Fermi energy fixed at $\varepsilon_{F}=0$, the dot energies at $\varepsilon=-5V$ and the interdot Coulomb interaction at $g=10V$; in the unit of the hybridization parameter $V$. In (a) We show the spin-spin correlation ${\bf \left\langle S_{1}\cdot S_{2}\right\rangle}$ between the dots for the half-filled case as a function of interdot tunneling matrix-element $t$ and ondot Coulomb interaction $U$. The possible values lie between $-\frac{3}{4}$ to $+\frac{1}{4}$. Positive values of ${\bf \left\langle S_{1}\cdot S_{2}\right\rangle}>0$ signifies the ferromagnetic correlation whereas the negative values ${\bf \left\langle S_{1}\cdot S_{2}\right\rangle}<0$ the antiferromagnetic correlation between the dots. (b) We show the average occupation of the dot-1 $\left\langle n_{1}\right\rangle$ as a function of interdot tunneling matrix-element $t$ and ondot Coulomb interaction $U$, can have values between $0$ to $2$. (c) We show the ground state of the half-filled system as a function of interdot tunneling matrix-element $t$ and ondot Coulomb interaction $U$. The integer value of $\eta$ identifies one of the possible ground states listed in Table \ref{tab:ksfe}. In the figure above, blue corresponds to $\eta =1$, yellow to $\eta =3$ and brown to $\eta =4$.
\label{fig:tUg10}\hfill \break}
\end{figure}
\\
\indent
For other higher values of the interdot Coulomb interaction $g$, the spin-spin correlation ${\bf \left\langle S_{1}\cdot S_{2}\right\rangle}$ between the dots exhibit similar behavior, as seen for $g=5$ in Fig. {\ref{fig:tUg5}}. For the sake of clarity, we also give the plot in Fig. {\ref{fig:tUg10}}, showing the behavior of spin-spin correlation in $U-t$ parameter space at a fixed value of interdot Coulomb interaction $g=10$. It is seen that the region corresponding to ferromagnetic correlation in $U-t$ parameter space in Fig. {\ref{corrg10}} is more than that for $g=5$ given in Fig. {\ref{corrg5}} but the value of spin-spin correlation ${\bf \left\langle S_{1}\cdot S_{2}\right\rangle}\gtrapprox 0$ is very small. This signifies that the spins on the dots are weakly coupled to form a triplet {\it i.e.} the ferromagnetic correlation is weak. This is due to the fact that the interdot Coulomb interaction $g$ causes the occupancies of the dots to be less than one $<n_{i}>\ < 1$ as can be seen from Fig. {\ref{avgg10}}. A triplet is formed when each of the dots contain an average of one electron $<n_{i}>\sim 1$. Thus as the value of interdot Coulomb interaction $g$ is increased further, the occupancies of the dots may go on decreasing. In Fig. {\ref{gsg10}}, the corresponding ground states $\left|\lambda_{\xi}^{4,\eta} \right\rangle$ in $U-t$ parameter space are given by the integer value $\eta$. It is seen that the ground states for $\eta=1$ and $\eta=3$ correspond to antiferromagnetic correlation and $\eta=4$ corresponds to ferromagnetic correlation between the dots.
\section{Conclusion}
\label{outcome}
The double quantum dot(DQD) system in parallel geometry with leads taken in the zero-bandwidth limit has been studied using exact diagonalization. The analytical forms of the eigenstates in each particle and spin sector with quantum numbers $\left(N,S,S_{z}\right)$ are obtained and the ground state in different regions of parameter space is identified from a four dimensional space in the half-filled system. It is observed that out of the four possible ground states listed in Table-{\ref{tab:gsfe}}, for a given set of parameters, the system can exist only in one of the three states $\left|\lambda_{\xi_{2}}^{4,1} \right\rangle$, $\left|\lambda_{\xi_{2}}^{4,3} \right\rangle$ and $\left|\lambda_{\xi_{2}}^{4,4} \right\rangle$. The spin-spin correlation between the dots is calculated for the ground state of the half-filled system. The model calculation shows that depending on the set of values of ondot Coulomb interaction $U$ and interdot tunneling matrix-element $t$, the spins at the two dots form either a singlet or a triplet. Even in the absence of interdot tunneling matrix-element $t$, the dots exhibit these two types of correlation through indirect exchange via the leads. The system parameters $(U,t)$ affect the occupancies of the dots in such a way that a large value of ondot Coulomb interaction {\it i.e.} $U\gg \left| \varepsilon \right|$ causes the occupancies of the dots to be restricted to $\left<n_{i}\right> \leqslant 1$ whereas the interdot tunneling matrix-element $t$ causes interdot charge transfer. 
It is the interplay of the above two effects that leads to different spin configurations of the dots.  
The ferromagnetic and antiferromagnetic configurations exhibit a sharp transition line in $(U,t)$ parameter space. This transition line is affected in the presence of interdot Coulomb interaction $g$. A very small value of interdot Coulomb interaction compared to the ondot Coulomb interaction $g\ll U$, leads to significant variation in the transition line. It is also observed that in the absence interactions, only antiferromagnetic correlation between the dots exist. Thus, a singlet or triplet state within DQDs in parallel geometry, can be probed when interactions are present in the system.   
\appendix
\section{ }
\label{Biquadsoln}
The values $m_{i}$ and $n_{i}$ used in different sections are given as
$m_{i}=\frac{1}{2}(q_{i}+k_{i}^2-r_{i}/k_{i})$, 
$n_{i}=\frac{1}{2}(q_{i}+k_{i}^2+r_{i}/k_{i})$ with 
$k_{i}^2=z_{i}-\frac{2}{3}q_{i}$, 
$z_{i}=\left( -{u_{i}}/{2}+w_{i}\right)^{\frac{1}{3}}+\left( -{u_{i}}/{2}-w_{i}\right)^{\frac{1}{3}}$, 
$w_{i}=\sqrt{{u_{i}^{2}}/{4}+{v_{i}^{3}}/{27}} $, 
$u_{i}=-\frac{2}{27}q_{i}^{3}+\frac{8}{3}p_{i}q_{i}-r_{i}^{2}$, 
$v_{i}=-4p_{i}-\frac{1}{3}q_{i}^{2}$,
$p_{i}=\mathtt{d_{i}}-\frac{3}{256}\mathtt{a_{i}}^4-\frac{1}{4}\mathtt{a_{i}}\mathtt{c_{i}}+\frac{1}{16}\mathtt{a_{i}}^2\mathtt{b_{i}}$, $q_{i}=\mathtt{b_{i}}-\frac{3}{8}\mathtt{a_{i}}^2$ and 
$r_{i}=\mathtt{c_{i}}-\frac{1}{2}\mathtt{a_{i}}\mathtt{b_{i}}+\frac{1}{8}\mathtt{a_{i}}^{3}$. 
\section{ } 
\label{shorthand}
The following short-hand notations have been used in various sections for writing basis states in different electron number sectors  
$\left|i\left(\bar{i}\right)\right\rangle \equiv c^{\dag}_{i\left(\bar{i}\right)\uparrow}c^{\dag}_{i\left(\bar{i}\right)\downarrow}\left|0\right\rangle$, 
$\left|l\left(\bar{l}\right)\right\rangle \equiv \alpha^{\dag}_{l\left(\bar{l}\right)\uparrow}\alpha^{\dag}_{l\left(\bar{l}\right)\downarrow}\left|0\right\rangle$, 
$\left|\sigma\left(\bar{\sigma}\right)\right\rangle \equiv c^{\dag}_{i\sigma\left(\bar{\sigma}\right)}c^{\dag}_{\bar{i}\sigma\left(\bar{\sigma}\right)}\left|0\right\rangle$, 
$\left|\Theta\right\rangle \equiv \frac{1}{\sqrt{2}}\left( c^{\dag}_{1\uparrow}c^{\dag}_{2\downarrow}+c^{\dag}_{1\downarrow}c^{\dag}_{2\uparrow}\right)\left|0\right\rangle$, 
$\left|\Xi\right\rangle \equiv \frac{1}{\sqrt{2}}\left( c^{\dag}_{1\uparrow}c^{\dag}_{2\downarrow}+c^{\dag}_{2\uparrow}c^{\dag}_{1\downarrow}\right)\left|0\right\rangle$, 
$\left|D\right\rangle \equiv c^{\dag}_{1\uparrow}c^{\dag}_{1\downarrow}c^{\dag}_{2\uparrow}c^{\dag}_{2\downarrow}\left|0\right\rangle$
 with $i\left(\bar{i}\right)=1\left(2\right)$, 
$l\left(\bar{l}\right)=s\left( a\right)$ and
$\sigma\left(\bar{\sigma}\right)=\uparrow\left(\downarrow\right)$.

\end{document}